\documentclass[fleqn,usenatbib]{mnras}

\usepackage{newtxtext,newtxmath}

\usepackage[T1]{fontenc}

\DeclareRobustCommand{\VAN}[3]{#2}
\let\VANthebibliography\thebibliography
\def\thebibliography{\DeclareRobustCommand{\VAN}[3]{##3}\VANthebibliography}

\usepackage{graphicx}	
\usepackage{amsmath}	
\usepackage{amssymb}	
\usepackage{bm}
\usepackage{caption}
\usepackage{multicol}
\usepackage{hyperref}
\usepackage{booktabs}
\usepackage{ulem}
\usepackage{mathtools}
\usepackage{lipsum}
\usepackage[most]{tcolorbox}
\RequirePackage{fix-cm}

\def\eqref#1{equation~(\ref{#1})}

\newcommand {\e}{\,\mathrm{e}}

\newcommand {\sech}{\,{\rm sech}}

\newcommand {\Myr}{\,{\rm Myr}}
\newcommand {\Gyr}{\,{\rm Gyr}}

\newcommand {\pc}{\,{\rm pc}}
\newcommand {\kpc}{\,{\rm kpc}}

\newcommand {\kpcGyr}{\,{\rm kpc}\,{\rm Gyr}^{-1}}
\newcommand {\kpcGyrkpcGyr}{\,{\rm kpc}^{2}\,{\rm Gyr}^{-2}}

\newcommand {\kpckpcGyr}{\,{\rm kpc}^2\,{\rm Gyr}^{-1}}

\newcommand {\phisun}{{\varphi_\odot}}

\newcommand {\Msun}{\,{\rm M}_\odot}

\newcommand {\vc}{v_{\rm c}}

\renewcommand{\deg}{{^{\circ}}}

\newcommand {\drm}{\mathrm{d}}

\newcommand {\vw}{{\bm w}}

\newcommand {\Rg}{R_{\rm g}}

\newcommand {\vz}{v_z}

\newcommand {\RCR}{R_{\rm CR}}

\newcommand {\zs}{z_{\rm s}}

\newcommand {\JR}{J_R}

\newcommand {\Jz}{J_z}

\newcommand {\Jphi}{J_\varphi}

\newcommand {\Js}{J_{\rm s}}

\newcommand {\Jsres}{J_{\rm s, res}}

\newcommand {\thetaz}{\theta_z}
\newcommand {\thetaphi}{\theta_\varphi}

\newcommand {\thetas}{\theta_{\rm s}}

\newcommand {\Omegap}{\Omega_{\rm p}}

\newcommand {\OmegaR}{\Omega_R}
\newcommand {\Omegar}{\Omega_r}
\newcommand {\Omegaz}{\Omega_z}

\newcommand {\Omegaphi}{\Omega_\varphi}

\newcommand {\Omegas}{\Omega_{\rm s}}

\newcommand {\Omegac}{\Omega_{\rm c}}

\newcommand {\Omegazres}{\Omega_{z,{\rm res}}}

\newcommand {\nz}{n_z}

\newcommand {\NR}{N_R}

\newcommand {\Nphi}{N_\varphi}
\newcommand {\Nz}{N_z}

\newcommand {\pd}{{\partial}}

\newcommand {\rhoc}{\rho_{\rm c}}

\newcommand {\td}{t_{\rm d}}

\title[]{Origin of the two-armed vertical phase-spiral in the inner Galactic disk}

\author[R. Chiba]{
Rimpei Chiba$^{1,2}$\thanks{E-mail: rimpei-chiba@g.ecc.u-tokyo.ac.jp},
Neige Frankel$^{2}$,
Chris Hamilton$^{3}$
\\
$^{1}$Department of Astronomy, Graduate School of Science, The University of Tokyo, 7-3-1 Hongo, Bunkyo-ku, Tokyo, 113-0033, Japan \\
$^{2}$Canadian Institute for Theoretical Astrophysics, University of Toronto, 60 St. George Street, Toronto, ON M5S 3H8, Canada \\
$^{3}$School of Natural Sciences, Institute for Advanced Study, Princeton, NJ 08540, USA \\
}

\date{Accepted XXX. Received YYY; in original form ZZZ}

\pubyear{2025}

\begin{document}
\label{firstpage}
\pagerange{\pageref{firstpage}--\pageref{lastpage}}
\maketitle

\begin{abstract}

\textit{Gaia} recently revealed a two-armed spiral pattern in the vertical phase-space distribution of the inner Galactic disk (guiding radius $\Rg \sim 6.2 \kpc$), indicating that some non-adiabatic perturbation symmetric about the mid-plane is driving the inner disk out of equilibrium. The non-axisymmetric structures in the disk (e.g., the bar or spiral arms) have been suspected to be the major source for such a perturbation. However, both the lifetime and the period of these internal perturbations are typically longer than the period at which stars oscillate vertically, implying that the perturbation is generally adiabatic. This issue is particularly pronounced in the inner Galaxy, where the vertical oscillation period is shorter and therefore adiabatically shielded more than the outer disk. We show that two-armed phase spirals can naturally form in the inner disk if there is a vertical resonance that breaks the adiabaticity; otherwise, their formation requires a perturber with an unrealistically short lifetime. We predict analytically and confirm with simulations that a \textit{steadily rotating} (non-winding) two-armed phase spiral forms near the resonance when stars are subject to both periodic perturbations (e.g., by spiral arms) and stochastic perturbations (e.g., by giant molecular clouds). Due to the presence of multiple resonances, the vertical phase-space exhibits several local phase spirals that rotate steadily at distinct frequencies, together forming a global phase spiral that evolves over time. Our results demonstrate that, contrary to earlier predictions, the formation of the two-armed phase spiral does not require transient perturbations with lifetimes shorter than the vertical oscillation period.

\end{abstract}




\section{Introduction}
\label{sec:introduction}

One of the most important discoveries made by the \textit{Gaia} satellite is arguably the remarkable spiral patterns found in the vertical phase-space of the Galactic stellar disk \citep{Antoja2018Nature}. These phase spirals are clear manifestations of phase mixing ongoing in the Milky Way. Phase mixing occurs because the period of the stars' vertical oscillation about the disk mid-plane increases with their amplitudes.

The phase spiral initially identified in the Solar neighborhood was predominantly one-armed. The one-armed phase spiral implies that stars in the disk have experienced a coherent kick that caused a dipole perturbation in the vertical phase-space distribution, which subsequently sheared into a spiral pattern. Possible sources of the perturbation include the Sagittarius dwarf galaxy \citep[e.g.,][]{Binney2018origin,Laporte2019Footprints,Li2020Dissecting,BlandHawthorn2021GalacticSeismology,Hunt2021Resolving,Asano2025Ripples}, the dark matter wake it induced \citep{Grand2023DarkMatterWake}, the buckling of the bar \citep{Khoperskov2019buckling}, bending waves in the disk \citep{Darling2019bending}, the cumulative effect of many weak perturbations from dark matter substructures \citep{Tremaine2023Snail,Gilman2024Dark}, and the `galactic echoes' from the nonlinear coupling of two-successive short-lived perturbations \citep{Chiba2025GalacticEchoes}.

\begin{figure}
  \begin{center}
    \includegraphics[width=8.4cm]{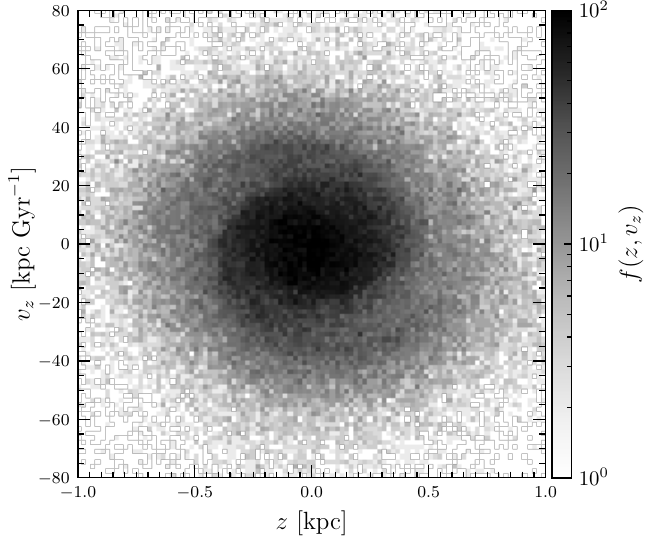}
    \includegraphics[width=8.4cm]{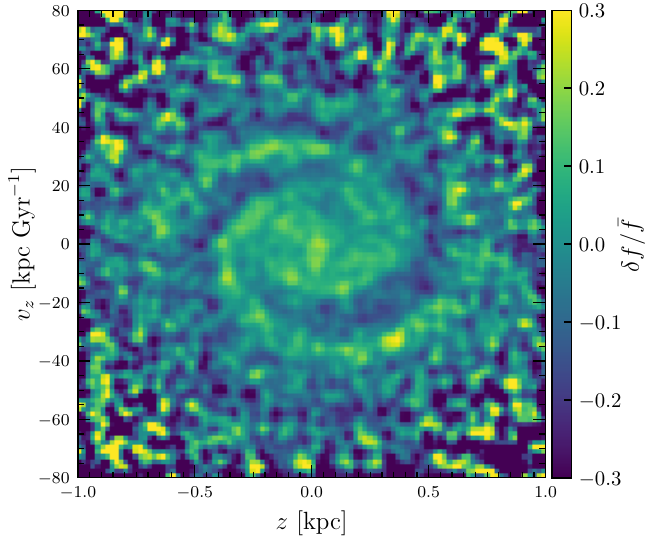}
    \caption{The two-armed phase-spiral discovered in \textit{Gaia} DR3. We select stars with parallax error $p/\sigma_p > 3$, magnitude $G<15$, cylindrical distance from the sun $d < 1\kpc$, angular momentum $\Jphi \in [1400,1600] \kpc^2 \Gyr^{-1}$, and azimuthal angle variable $|\thetaphi - \phisun| < 5 \deg$. The angle variables are computed using the Milky Way potential from McMillan (2017) and the St\"ackel fudge \citep{Binney2012Stackel} as implemented in AGAMA \citep{Vasiliev2019AGAMA}. Top panel shows the full distribution, while the bottom panel shows the fractional density contrast relative to the smooth distribution obtained using a Gaussian filter with scale $80 \pc$ in $z$ and $6.4 \kpcGyr$ in $v_z$.}
    \label{fig:gaiaDR3_zvz_df}
  \end{center}
\end{figure}

Recently, \cite{Hunt2022Multiple} mapped the phase spiral across the disk using the \textit{Gaia}'s third data release (DR3), and discovered a two-armed phase spiral in the inner disk (Fig.\,\ref{fig:gaiaDR3_zvz_df}).\footnote{We refer to stars in the `inner disk' as those with small guiding radius ($\Rg \sim 6.2 \kpc$) and not necessary those located in the inner disk. The stars analysed in \cite{Hunt2022Multiple} are in fact all restricted to the Solar neighborhood.} This finding came as a surprise, since both analytical theories \citep{Widrow2014Bending,Banik2022ComprehensiveI,Banik2023ComprehensiveII} and $N$-body simulations \citep{Hunt2021Resolving} predicted that encounters with satellite galaxies will produce two-armed phase spirals in the outer disk rather than the inner disk due to tidal forces. To date, the two-armed phase spiral has been detected in various quantities, including the mean radial velocity \citep{Li2023Gaia} and the mean metallicity \citep{Alinder2024Limitations}, both of which are less affected by selection effects.

\cite{Hunt2022Multiple} demonstrated using high-resolution $N$-body simulations that two-armed phase spirals can be generated in the inner disk by the disk's internal perturbations, such as the galactic bar and spiral arms. More recently, \cite{Asano2025Ripples} confirmed their finding with an even higher-resolution simulation and proposed a scenario in which the two-armed phase spirals were created by spiral arms tidally induced by the impact of the Sagittarius dwarf galaxy. These simulations demonstrate a clear causal link between the two-armed phase spiral and the internal disk perturbation. However, the underlying physical mechanism remains puzzling. While it has been known that bars and spiral arms induce breathing motions in the disk \citep[e.g.,][]{Debattista2014vertical,Faure2014Radial,Monari2015vertical,Monari2016Modelling,Khachaturyants2022pattern,Kumar2022Excitation,Asano2024Growing}, it is unclear how they could generate phase spirals, which specifically require a \textit{non-adiabatic} perturbation to the disk, i.e., the timescale of the perturbation must be short compared to the vertical oscillation period -- otherwise, the star's vertical action would be conserved. \cite{Li2023Gaia} provide an example of a disk perturbed by a steadily rotating bar and a spiral arm which exhibits a breathing motion but no phase spirals.\footnote{Note that \cite{Li2023Gaia} also show that a decelerating bar with sweeping resonances can temporarily generate a pronounced two-armed phase spiral, highlighting the key role of resonances in the formation of such structures (see discussion in Section \ref{sec:galactic_bar}).} \cite{Banik2022ComprehensiveI,Banik2023ComprehensiveII} performed linear analyses of the disk response to various perturbations and suggested that the formation of a two-armed phase spiral requires the internal perturber to rapidly grow and decay over a timescale comparable to the vertical oscillation period, which can be as short as $70 \Myr$ in the inner disk.

This paper provides an explanation as to why two-armed phase spirals can form in the inner disk despite its very short dynamical period, using a combination of linear analysis, nonlinear analysis, and test-particle simulations. The key ingredients are resonance and diffusion. It is well known that a resonance can break the adiabaticity of the system even in a slowly varying potential (see \citealt{Weinberg1994Adiabatic} for a general discussion). However, while a resonant perturbation can temporarily generate an open phase spiral, it alone will eventually yield a closed ring-like structure composed of a chain of resonant islands (trapped phase-space), as stars near resonance librate and gradually phase mix within the trapped region. When these stars are also subject to small-scale stochastic kicks, which inevitably exist in both real and simulated galaxies, the phase mixing along the motion of libration remains incomplete, allowing the open spiral pattern to be preserved. As we elaborate in the main text, this competition between persistent resonant forcing and diffusion due to small-scale kicks results in an open two-armed phase spiral that rotates \textit{steadily} without winding, with its shape set by the diffusion timescale rather than the timing of the perturbation, similar to the argument by \cite{Tremaine2023Snail}. In the case of a transient perturbation, the phase spiral starts winding up once the perturbation has decayed.

The structure of the paper is as follows: We first lay out our model in Section \ref{sec:model}. In Section \ref{sec:linear_theory}, we solve the linearized kinetic equation and analytically predict the emergence of a steady two-armed phase-spiral. In Section \ref{sec:nonlinear_theory}, we provide a complementary, more intuitive explanation to the formation mechanism of the phase spiral using a non-linear approach. In Section \ref{sec:simulation}, we study the evolution of the phase spiral using test-particle simulations. Section \ref{sec:discussion} discusses the origin of the two-armed phase spiral in the Milky Way and Section \ref{sec:summary} concludes our study.

\section{Model}
\label{sec:model}

To study the formation mechanism of the two-armed phase spiral with minimal complexity, we restrict our model to one dimension in the direction normal to the galactic mid-plane $z$. We further ignore the self-gravity of the perturbation, which can alter the amplitude and winding rate of the phase spiral \citep{Darling2019bending,Widrow2023Swing}.

\subsection{Isothermal slab}
\label{sec:Isothermal_slab}

\begin{figure}
  \begin{center}
    \includegraphics[width=8.4cm]{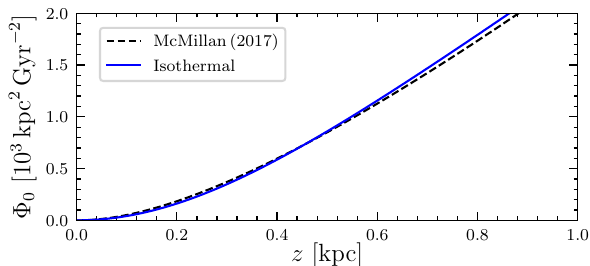}
    \includegraphics[width=8.4cm]{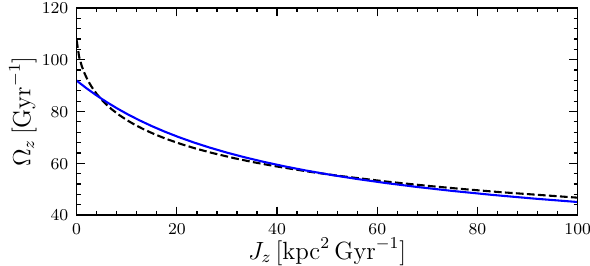}
    \caption{Top: Potential of the isothermal slab fitted to the Milky Way potential from McMillan (2017) at $R=6.2 \kpc$. Bottom: Vertical orbital frequency as a function of the vertical action for stars with $(\JR,\Jphi)=(0,R\vc)$.}
    \label{fig:Phi}
  \end{center}
\end{figure}

Following \cite{Banik2022ComprehensiveI} and \cite{Tremaine2023Snail}, we model the unperturbed galactic disk with a self-gravitating isothermal slab:
\begin{align}
  \Phi_0(z) = 2 \sigma^2 \ln \left[ \cosh \left(\frac{z}{2h}\right)\right], ~~~~
  \rho_0(z) = \rhoc \sech^2 \left(\frac{z}{2h}\right),
  \label{eq:isothermal_slab_Phi0_rho0}
\end{align}
where $h$ is the scale height at large $z$, $\sigma$ is the velocity dispersion, and $\rhoc = \sigma^2 / (8\pi G h^2)$. We adopt $h=0.2 \kpc$ and $\sigma = 26 \kpc \Gyr^{-1}$, which give a reasonable approximation to the vertical potential of the Milky Way \citep{McMillan17} at $R=6.2 \kpc$ as shown in the top panel of Fig.\,\ref{fig:Phi}. The distribution function of the isothermal slab is 
\begin{align}
  f_0(z,v_z)  &= \frac{\rhoc}{(2 \pi \sigma^2)^{1/2}} \exp\left[-\frac{H_0(z,v_z)}{\sigma^2}\right],
  \label{eq:isothermal_slab_f0}
\end{align}
where $H_0(z,v_z) = v_z^2/2 + \Phi_0(z)$ is the unperturbed Hamiltonian. The set of functions (\ref{eq:isothermal_slab_Phi0_rho0})-(\ref{eq:isothermal_slab_f0}) satisfy
\begin{align}
  \nabla^2 \Phi_0(z) = 4\pi G \rho_0(z) ~~ {\rm and } ~~ \rho_0(z) = \int \drm v_z f_0(z,v_z).
  \label{eq:Poisson_eq}
\end{align}
Since $H_0$ is integrable, we may define angle-action variables $(\thetaz,\Jz)$: 
\begin{align}
  \thetaz = \Omega_z(\Jz) \int_0^z \frac{\drm z}{v_z}, ~~~~ \Jz = \frac{1}{2\pi} \oint \drm z ~v_z,
  \label{eq:angle_action}
\end{align}
where $\Omega_z(\Jz)$ is the orbital frequency
\begin{align}
  \Omega_z(\Jz) = 2 \pi \left( \oint \frac{\drm z}{v_z} \right)^{-1},
  \label{eq:frequency}
\end{align}
which reduces to the vertical epicycle frequency in the limit $\Jz \rightarrow 0$
\begin{align}
  \nu = \frac{\sigma}{\sqrt{2}h}.
  \label{eq:epicycle_frequency}
\end{align}
The bottom panel of Fig.\,\ref{fig:Phi} plots $\Omega_z$ as a function of $\Jz$. Compared to the Milky way model, the isothermal potential has a smaller gradient $\drm \Omegaz/\drm \Jz$ at small $\Jz$ due to less mass near the mid-plane, implying a lower rate of phase mixing in that region.

For subsequent analysis, we also define the following Cartesian canonical coordinates:
\begin{equation}
  \begin{aligned}
    q &= \sqrt{2\Jz} \sin \thetaz \\
    p &= \sqrt{2\Jz} \cos \thetaz
  \end{aligned}
  ~~~
  \xleftrightarrow{\quad}
  ~~~
  \begin{aligned}
    \Jz &= (q^2 + p^2)/2 \\
    \thetaz &= \arctan(q/p).
  \end{aligned}
  \label{eq:cartesian}
\end{equation}

\subsection{Periodic spiral perturbation}
\label{sec:model_periodic_perturbation}

\begin{figure*}
  \begin{center}
    \includegraphics[width=17cm]{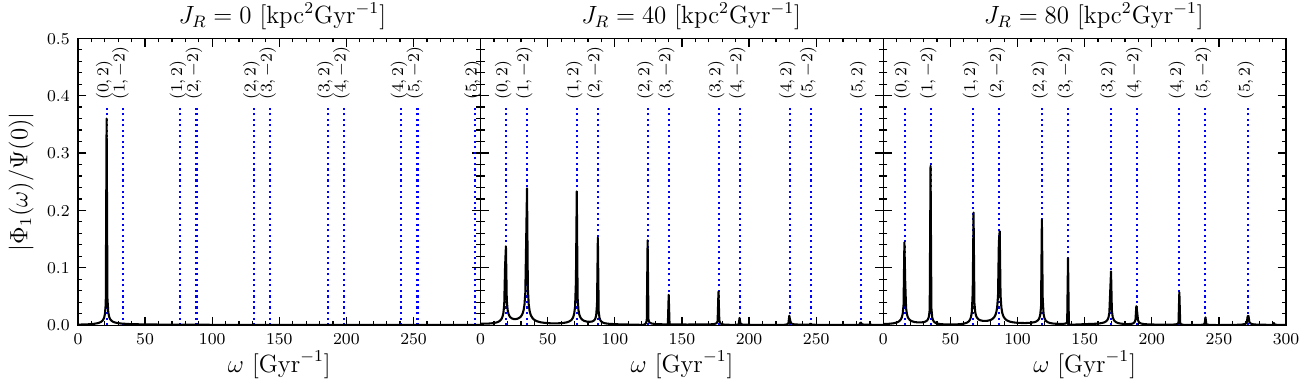}
    \caption{Spectra of the spiral arms' potential along an in-plane unperturbed orbit with guiding radius $\Rg=6.2\kpc$ for three different values of the radial action $\JR$. The brackets in the figures denote the set of integers $(\NR,\Nphi)$, corresponding to the frequency $\omega=\NR\OmegaR+\Nphi(\Omegaphi-\Omegap)$. The two-armed phase spiral was observed from stars with large radial actions ($\JR\simeq80\kpckpcGyr$, right panel). These stars are subject to a variety of resonances at high frequencies.}
    \label{fig:spectra}
  \end{center}
\end{figure*}

We perturb the isothermal slab with a spiral arm, which generates a potential perturbation that is symmetric about the galactic mid-plane. A similar role may be played by the Galactic bar, although it is less likely to be responsible for the observed two-armed phase spiral for reasons we discuss in Section \ref{sec:galactic_bar} and Appendix \ref{sec:app_bar_resonance}.

We model the potential of the spiral arm with the following function:
\begin{align}
  \Phi_1(z,t) &= \Psi(z) \mathcal{T}(t) \cos (\omega t),
  \label{eq:spiral_potential}
\end{align}
where $\Psi(z)$ describes the potential's vertical profile and is an even function of $z$. The potential of a tightly wound spiral arm in a razor thin disk is known to decay exponentially with $z$, i.e., $\Psi(z) \propto \e^{-k|z|}$ \citep{binney2008galactic}, where $k$ is the radial wave number of the spiral. The resulting force discontinuously switches sign at the mid-plane. To avoid such discontinuity, we adopt the following softened potential 
\begin{align}
  \Psi(z) &= - \frac{2\pi G \Sigma_{\rm max} }{k} a K_1(a) \e^{- k \sqrt{z^2+z_{\rm s}^2}},
  \label{eq:spiral_potential_vertical}
\end{align}
where $z_{\rm s}$ is the softening length, $a = k z_{\rm s}$, and $K_1$ is the modified Bessel function of the second kind (Appendix \ref{sec:app_spiral_model}). With this form, the amplitude of the vertically-integrated surface density $\Sigma_{\rm max}$ is equivalent to that of spiral arms in a razor-thin disk. This eases comparison with other studies employing a razor-thin model, in particular, the recent measurement $\Sigma_{\rm max} = 5.5 \Msun \pc^{-2}$ by \cite{Eilers2020Spiral}. By default, we set $z_{\rm s} = 0.1 \kpc$ and $k=1.5 \kpc^{-1}$. The latter corresponds to the radial wave number of an $m=2$ armed logarithmic spiral ($k=m\cot\alpha/R$) at $R=6.2 \kpc$ with pitch angle $\alpha=12^\circ$ as measured in the Milky Way \citep{Vallee2015Different,Eilers2020Spiral}.

The function $\mathcal{T}(t)$ in equation (\ref{eq:spiral_potential}) describes the time variation of the spiral amplitude. As noted by \cite{Banik2023ComprehensiveII}, the response of the disk to the spiral perturbation varies depending on whether the spiral arm is short-lived (`transient') or long-lived (`persistent'). We thus consider the following two models:
\begin{align}
  \mathcal{T}(t) = 
  \begin{cases}
    \e^{-(t - t_{\rm p})^2 / (2 \sigma_t^2)} \hspace{17mm} {\rm (transient)}, \\
    \e^{-(t - t_{\rm p})^2 \Theta(t_{\rm p} - t) /(2 \sigma_t^2)} \hspace{8mm} {\rm (persistent)},
  \end{cases}
  \label{eq:spiral_potential_temporal}
\end{align}
where $t_{\rm p}$ is the time at which the amplitude reaches its peak, $\sigma_t$ is the characteristic growth/decay time of the spiral arm\footnote{Note that the total lifetime of a transient spiral arm with a Gaussian profile can extend arbitrarily beyond $\sigma_t$. For reference, the full-width-at-tenth-maximum is $2 \sqrt{2\ln 10} \sigma_t \sim 4.29 \sigma_t$.}, and $\Theta$ is a Heaviside step-function. The transient model grows and decays as a Gaussian, while the persistent model grows up to $t_{\rm p}$ and stays constant thereafter. 

Lastly, we need to determine the perturbing frequency $\omega$, which sets the location of the resonance. In the simplest case where all stars follow circular orbits, $\omega=m(\Omegac - \Omegap)$, where $m$ denotes the number of spiral arms, $\Omegac$ the circular frequency, and $\Omegap$ the pattern speed. For $m=2$, $\Omegac = 38 \Gyr^{-1}$ at $\Rg=6.2 \kpc$, and $\Omegap = 28 \Gyr^{-1}$ \citep{Grosbol2018spiral,Dias2019spiral,Monteiro2021distribution}, $\omega$ would be $20 \Gyr^{-1}$, which is much smaller than the typical vertical frequency $\Omegaz = 50-100 \Gyr^{-1}$ at $\Rg=6.2\kpc$ (see Fig.\,\ref{fig:Phi}). This suggests that a direct resonance between the vertical and azimuthal motion of stars is absent in the vertical phase-space of interest. Crucially, however, the two-armed phase spiral is detected from stars on rather eccentric orbits, with guiding radii well inside that of the sun: stars with small guiding radii need large radial motion to reach the solar neighbourhood. Local stars with guiding radius $\Rg \simeq 6.2 \kpc$ typically have $\JR \simeq 80 \kpckpcGyr$, corresponding to an epicycle radius of $\delta R \simeq 2 \kpc$. Along such a highly non-circular orbit, the time variation of the spiral's potential becomes more complicated with high frequencies involved.

Fig.\,\ref{fig:spectra} shows the spectra of the spiral perturbation along an unperturbed orbit with $\Jz=0$ (in-plane), $\Jphi=1488 \kpckpcGyr$ (guiding radius $\Rg = 6.2 \kpc$), and three different radial actions $\JR$. We assumed a flat rotation curve $\vc=240\kpc\Gyr^{-1}$ \citep{Eilers2019Circular,Mroz2019Rotation,Reid2019Trigonometric,Ablimit2020Rotation} and a logarithmic spiral arm with a Gaussian radial profile in surface density that peaks at the corotation radius.\footnote{$N$-body simulations typically find that the surface density of spiral arms have peak amplitude near their corotation radius \citep[e.g.,][]{Sellwood2011lifetimes,Wada2011Interplay,Grand2012Spiral,Baba2013Dynamics,Kawata2014Gas,VeraCiro2014Effect}, and the co-existence of a number of such spiral arms give rise to an apparently shearing, transient pattern \citep{SellwoodCarlberg2014,Sellwood2019SpiralInstabilities,Sellwood2021Spiral}. This is supported by observations of the spiral's pattern speed in the Milky Way \citep{CastroGinard2021Milky,Joshi2023Revisiting} and the variation of the vertex deviation across the spiral arms \citep{Funakoshi2024Clues}. A radial profile declining away from corotation is also necessary to reduce radial heating at the Lindblad resonances \citep{Hamilton2024Why} and keep the disk cold \citep{Aumer2016Age,Frankel2020Keeping}.} The corresponding potential perturbation is (Appendix \ref{sec:app_spiral_model})
\begin{align}
  \Phi_1(R,\varphi,t) =& - \frac{2 \pi G \Sigma_{\rm max}}{m \cot\alpha} a K_1(a) R \e^{-(R-\RCR)^2/(2R_\beta^2)} \nonumber \\
   & \times \cos \left[ m (\varphi - \Omegap t + \cot \alpha \ln R) \right],
  \label{eq:spiral_potential_2d}
\end{align}
where $R_\beta \equiv \beta \sqrt{2}\RCR / m$ is the Gaussian width with the parameter $\beta$ describing the ratio between $R_\beta$ and the distance between the corotation and Lindblad resonances \citep{Hamilton2024Why}. We set $m=2$, $\beta=0.5$ \citep{Hamilton2024Why}, pitch angle $\alpha=12^\circ$ \citep{Vallee2015Different,Eilers2020Spiral}, and pattern speed $\Omegap=28\Gyr^{-1}$ \citep{Grosbol2018spiral,Dias2019spiral,Monteiro2021distribution}, corresponding to the Sagittarius-Carina arm \citep{Naoz2007Open,CastroGinard2021Milky,Joshi2023Revisiting}. Along a circular orbit $\JR=0$ (Fig.\,\ref{fig:spectra}, left plot), the spiral's potential oscillates at a single frequency $\omega=m(\Omegac - \Omegap) \sim 20 \Gyr^{-1}$. As the radial action of stars increases from left to right, the spectrum spreads significantly to large $\omega$ as stars can now enter and leave the spiral arm due to their epicycle motion. The numerous peaks correspond to integer multiples of the radial and azimuthal frequencies $(\OmegaR,\Omegaphi)$:
\begin{align}
  \omega=\NR\OmegaR+\Nphi(\Omegaphi-\Omegap), 
  \label{eq:resonance}
\end{align}
where the integers $(\NR,\Nphi)$ are denoted in the figures. A resonance occurs when $\omega$ is in commensurable relation with the vertical frequency, i.e., $\omega = \Nz\Omegaz$. Fig.\,\ref{fig:spectra} suggests that the large radial motion of the low-$\Jphi$ stars observed in the Solar neighborhood allows multiple resonances to lie at the relevant vertical phase-space, $\Omegaz = 50-100 \Gyr^{-1}$ (or $\omega = 100-200 \Gyr^{-1}$).

As we will see in Section \ref{sec:simulation_Jr80}, the phase-space dynamics at each resonance is fundamentally the same. Hence, for most of our following analysis, we will focus on the disk response around a single resonance (single $\omega$). By default, we set $\Omegazres=0.9\nu$ $(\omega = 1.8 \nu \sim 165 \Gyr^{-1})$, which, for example, corresponds to the $(\NR,\Nphi,\Nz)=(3,2,2)$ resonance in our fiducial spiral-arm model with $\Omegap=28 \Gyr^{-1}$. We will explore the dependence of the response on $\omega$ in Section \ref{sec:persistent_spiral_arm}. Later, in Section \ref{sec:simulation_Jr80}, we will extend our analysis to include multiple resonances for various $\Omegap$ and discuss the possible resonances that could be responsible for the observed phase spiral.

\subsection{Stochastic perturbation}
\label{sec:model_stochastic_perturbation}

In addition to the periodic perturbation by spiral arms, we subject the disk stars with small-scale stochastic kicks due to gravitational encounters with, for example, giant molecular clouds \citep[e.g.,][]{Carlberg1987Vertical,Jenkins1990Spiral,Aumer2016Age} and dark matter substructures \citep[e.g.,][]{Toth1992Galactic,Kazantzidis2009Cold}. Following \cite{Tremaine2023Snail}, we model this process by a Gaussian random walk in $(q,p)$ space, that is, the \textit{ensemble average} of the displacement is zero, $\langle \Delta q \rangle = \langle \Delta p \rangle = 0$, while that of the squared displacement over time $\Delta t$ is
\begin{align}
  \langle (\Delta q)^2 \rangle = D \Delta t, ~~~ \langle (\Delta p)^2 \rangle = D \Delta t,
  \label{eq:diffusion}
\end{align}
where $D$ is the diffusion coefficient, which we assume to be constant. Under this process, it can be shown that the mean action of the disk changes at rate, $\drm \overline{\Jz}/ \drm t = D$ (Appendix \ref{sec:app_diffusion}). To determine the diffusion coefficient, we assume that the disk was initially razor thin (i.e., its initial mean action is zero, $\overline{\Jz}=0$) and was gradually heated over time $T$ to reach its present mean action, $\overline{\Jz}(T)$. Then
\begin{align}
  D = \frac{\overline{\Jz}(T)}{T} = \frac{1}{T} \int \drm \theta_z \int \drm \Jz \Jz f_0(\Jz) \sim \frac{1.80 \sigma h}{T}.
  \label{eq:diffusionoefficient}
\end{align}
By default, we adopt $T=10 \Gyr$, which gives $D \simeq 0.94 \kpcGyrkpcGyr$.

We note that subjecting stars to random walk in the $(q,p)$ space or $(z,v_z)$ space results in little difference (Appendix \ref{sec:app_diffusion}), provided we determine the diffusion coefficients in $(z,v_z)$ from their present mean-squared value
\begin{align}
  D_z = \frac{\overline{z^2}}{T} \sim \frac{3.62 h^2}{T}, ~~~~ 
  D_v = \frac{\overline{v_z^2}}{T} = \frac{\sigma^2}{T}.
  \label{eq:diffusionoefficient_zvz}
\end{align}
Therefore, in our test-particle simulations (Section \ref{sec:simulation}), we apply kicks in the $(z,v_z)$ coordinates, which is computationally less expensive.

\section{Linear analysis}
\label{sec:linear_theory}

In this section, we study the disk response to spiral arms under stochastic perturbations using linear theory. The general framework has been laid out in detail by \cite{Banik2023ComprehensiveII}. Here, we examine the particular case where the disk is subject to \textit{persistent} spiral arms in the presence of small-scale random kicks, which was not analysed in \cite{Banik2023ComprehensiveII}. As we shall see, this leads to an intriguing result that a steady (non-winding) spiral pattern forms in phase space. 

The evolution of the distribution function (DF) $f$ with a stochastic process is governed by the kinetic equation:
\begin{align}
  \frac{\pd f}{\pd t} + [f,H] = C[f],
  \label{eq:kinetic_eq}
\end{align}
where $H$ is the Hamiltonian of the system, $[\cdot,\cdot]$ is the Poisson bracket, and $C[f]$ is the `collision' operator. The Gaussian random walk introduced in Section \ref{sec:model_stochastic_perturbation} is described by the following collision operator (Appendix \ref{sec:app_diffusion})
\begin{align}
  C[f] &= \frac{1}{2} D\left(\frac{\pd^2 f}{\pd q^2} + \frac{\pd^2 f}{\pd p^2} \right)
  = D\left(\frac{\pd}{\pd \Jz} \Jz \frac{\pd f}{\pd \Jz} + \frac{1}{4\Jz}\frac{\pd^2 f}{\pd \thetaz^2} \right).
  \label{eq:collision_operator}
\end{align}
As we will find, the perturbation by a persistent spiral arm is localized near resonances, so derivatives with respect to $\Jz$ dominate. Hence we may approximate
\begin{align}
  C[f] &\simeq D \Jz \frac{\pd^2 f}{\pd \Jz^2}.
  \label{eq:collision_operator_approx}
\end{align}
We expand the Hamiltonian $H=H_0(\Jz)+\Phi_1(\thetaz,\Jz,t)$ and the DF $f=f_0(\Jz)+f_1(\thetaz,\Jz,t)$, where $f_0$ is the initial unperturbed DF. The linearized kinetic equation is then 
\begin{align}
  \frac{\pd f_1}{\pd t} + [f_1,H_0] + [f_0,\Phi_1] = C[f_0] + C[f_1].
  \label{eq:kinetic_eq_lin}
\end{align}
Fourier transforming the perturbations in $\thetaz$,
\begin{align}
  f_1(\thetaz,\Jz,t) =& \sum_{\nz} \hat{f}_{\nz}(\Jz,t) \e^{i \nz \thetaz}, \\
  \Phi_1(\thetaz,\Jz,t) =& \sum_{\nz} \hat{\Phi}_{\nz}(\Jz,t) \e^{i \nz \thetaz},
  \label{eq:Fourier_expand_f1_Phi1}
\end{align}
we have
\begin{align}
  \frac{\pd \hat{f}_{\nz}}{\pd t} + i \nz \Omegaz \hat{f}_{\nz} - i \nz \frac{\pd f_0}{\pd \Jz} \hat{\Phi}_{\nz} = D \Jz \frac{\pd^2}{\pd \Jz^2} \left(\delta^0_{\nz} f_0 + \hat{f}_{\nz}\right).
  \label{eq:kinetic_eq_lin_Fourier}
\end{align}
The equation for $\nz=0$ describes the slow evolution of the angle-independent distribution due to diffusion. Here, we are interested in the evolution of the angle-dependent perturbation ($\nz\neq0$), for which we have the following approximate solution, given the initial condition $\hat{f}_{\nz}=0$ (Appendix \ref{sec:app_solution_LKE}):
\begin{align}
  \hat{f}_{\nz}(\Jz,t) \!=\! i \nz \frac{\pd f_0}{\pd \Jz} \!\int_0^t \!\drm t' \e^{-i \nz \Omegaz (t-t') - \left[(t-t')/\td\right]^3} \hat{\Phi}_{\nz}(\Jz,t'),
  \label{eq:kinetic_eq_lin_sol_fn}
\end{align}
where
\begin{align}
  \td \equiv [3/(\nz^2 \Omegaz'^2 \Jz D)]^{1/3}
  \label{eq:diffusion_timescale}
\end{align}
is the diffusion timescale. The solution contains a super-exponential decay factor, $\exp \left[-(t/\td)^3\right]$, arising from the joint effect of phase mixing and diffusion: phase mixing turns structures into increasingly smaller scale, while diffusion efficiently erases the resulting small-scale structures. Without phase mixing, perturbations periodic in $\Jz$ will undergo an exponential decay, which is much slower than super-exponential.

The Fourier coefficient of the spiral perturbation (equation \ref{eq:spiral_potential}) is
\begin{align}
  \hat{\Phi}_{\nz}(\Jz,t) = \hat{\Psi}_{\nz}(\Jz) \mathcal{T}(t) \sum_{l = \pm 1} \frac{1}{2} \e^{- i l \omega t},
  \label{eq:spiral_potential_Fourier_coefficient}
\end{align}
where 
\begin{align}
  \hat{\Psi}_{\nz}(\Jz) \equiv \frac{1}{2\pi} \int \drm \thetaz \Psi(z) \e^{- i \nz \thetaz}.
  \label{eq:Zk}
\end{align}
Substituting (\ref{eq:spiral_potential_Fourier_coefficient}) to (\ref{eq:kinetic_eq_lin_sol_fn}), switching variable to $\tau = t - t'$, and inverse Fourier transforming, we obtain the angle-dependent part of $f_1$:
\begin{align}
  f_1(\thetaz,\Jz,t) =& \frac{1}{2} \sum_{\nz, l =\pm 1} i \nz \frac{\pd f_0}{\pd \Jz} \hat{\Psi}_{\nz} \e^{- i l \omega t} \e^{i \nz \thetaz} \nonumber \\
  &\times \int_0^t \drm \tau \, \mathcal{T}(t - \tau) \e^{-i \Omegas \tau - \left(\tau/\td\right)^3},
  \label{eq:kinetic_eq_lin_sol}
\end{align}
where $\Omegas = \nz \Omegaz - l \omega$ is the `slow' frequency, which is zero at the resonance, $(\nz,l)=(\pm \Nz, \pm 1)$. In the time-asymptotic limit $(t \rightarrow \infty)$, the integral converges to a constant value since $\mathcal{T} \rightarrow 1$ and $\exp[-(\tau/t_{\rm d})^3] \rightarrow 0$. This leaves the time dependence of equation (\ref{eq:kinetic_eq_lin_sol}) solely in the exponential function, $\exp(- i l \omega t)$, indicating that long after the perturbation has emerged, the response rotates steadily in phase space at the forcing frequency $\omega$.

\begin{figure}
  \begin{center}
    \includegraphics[width=8.4cm]{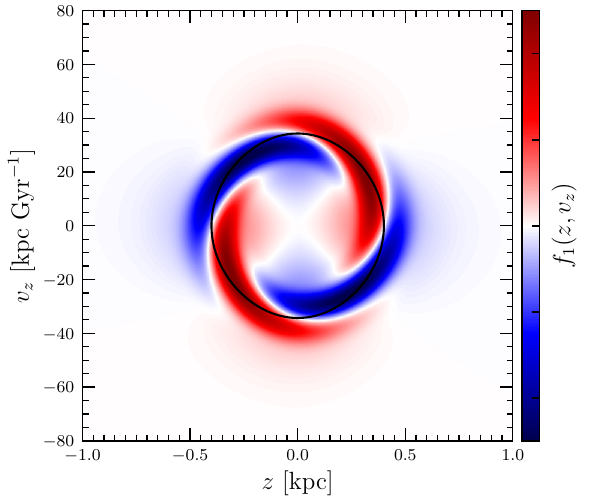}
    \includegraphics[width=8.4cm]{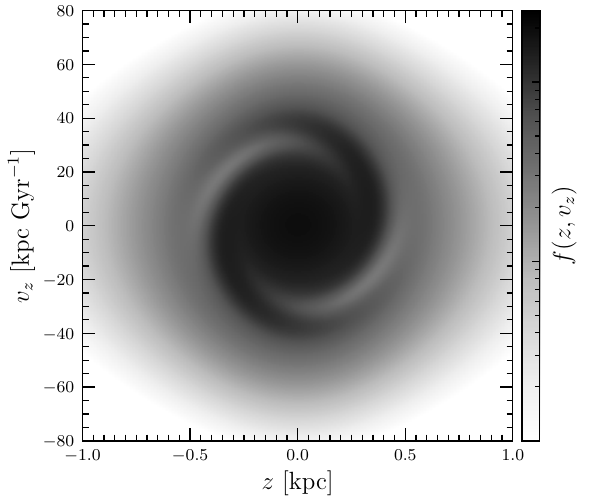}
    \caption{A steadily rotating (non-winding) two-armed phase-spiral predicted by the linearized kinetic equation (\ref{eq:kinetic_eq_lin_sol}) in the time-asymptotic limit. The disk is subject to a periodic vertical perturbation by galactic spiral arms in the presence of small-scale random kicks. The black curve marks the resonance.}
    \label{fig:linear_zvz}
  \end{center}
\end{figure}

We now examine the shape of the response in the time-asymptotic limit. Let us set $\mathcal{T}=1$ (ignore the transient effects due to the growth of the spiral arm) and solve the limit $t \rightarrow \infty$ of the integral by Taylor expanding the integrand in $\Omegas$ about the resonance $\Omegas = 0$:
\begin{align}
   \int_{0}^{\infty} \drm \tau \e^{-i \Omegas \tau - \left(\tau/\td\right)^3}
   &= \int_{0}^{\infty} \drm \tau \sum_{k=0} \frac{(-i \Omegas \tau)^k}{k!} \e^{- \left(\tau/\td\right)^3} \nonumber \\
   &= \frac{\td}{3} \sum_{k=0} \frac{(-i \Omegas \td)^k}{k!} \Gamma \left(\frac{k+1}{3}\right),
  \label{eq:integral_expand}
\end{align}
where $\Gamma$ is the gamma function. Since the response is localized at the resonance, we truncate the series up to $k=1$:
\begin{align}
  \int_{0}^{\infty} \drm \tau \e^{-i \Omegas \tau - \left(\tau/\td\right)^3}
  &\simeq \frac{\td}{3} \left[ \Gamma\left(\frac{1}{3}\right) -i \Omegas \td \Gamma\left(\frac{2}{3}\right) + \mathcal{O}(\Omegas^2\td^2)\right] \nonumber \\
  &\simeq \frac{\td}{3} \Gamma\left(\frac{1}{3}\right) \e^{- i \alpha \Omegas \td},
  \label{eq:integral_expand}
\end{align}
where $\alpha \equiv \Gamma(2/3)/\Gamma(1/3) \simeq 0.51$. The diffusion timescale at the resonance is $\td \simeq 0.35 \Gyr$, so the above truncation is valid only in the close vicinity of the resonance $|\Omegas| < 1/\td \simeq 3 \Gyr^{-1}$. Plugging the result back to equation (\ref{eq:kinetic_eq_lin_sol}), we have
\begin{align}
  f_1(\thetaz,\Jz,t) \simeq \frac{1}{2} \sum_{\nz, l = \pm 1} i \nz \frac{\pd f_0}{\pd \Jz} \hat{\Psi}_{\nz} \e^{- i l \omega t} \frac{\td}{3} \Gamma\left(\frac{1}{3}\right) \e^{i (\nz \thetaz - \alpha \Omegas \td)}. 
  \label{eq:kinetic_eq_lin_sol_approx}
\end{align}
The factor, $\exp[i(\nz \thetaz - \alpha \Omegas \td)]$, implies that the shape of the steadily rotating response winds with increasing distance from the resonance, indicating a steady phase spiral. The phase spiral is predicted to be trailing consistent with the observed pattern (Fig.\,\ref{fig:gaiaDR3_zvz_df}). The degree of winding is set by the diffusion time $\td$, similar to the prediction by \cite{Tremaine2023Snail}, who explored the origin of the one-armed phase spiral in the presence of diffusion. The equation also exhibits a factor, $\td$, which implies that the response vanishes in the strongly diffusive limit ($\td \rightarrow 0$) as expected.

To explicitly demonstrate the formation of a steadily rotating phase-spiral, we show in Fig.\,\ref{fig:linear_zvz} the disk response to persistent spiral arms calculated using equation (\ref{eq:kinetic_eq_lin_sol}), where we performed the integral in $\tau$ numerically. The gradient of the unperturbed distribution function, $f'_0(\Jz)$, the Fourier amplitude of the perturbation, $\hat{\Psi}_{\nz}(\Jz)$, and the diffusion timescale $t_{\rm d}(\Jz)$ are all calculated according to the models presented in Section \ref{sec:model}. The upper panel shows the perturbation $f_1$, while the lower panel shows the full distribution $f=f_0+f_1$. As predicted, a trailing two-armed phase spiral forms near the resonance marked by the black curve. Over time, this phase spiral rotates at a constant rate due to the factor, $\exp(-i l \omega t)$. Here, we have chosen an arbitrary phase. Despite all the simplified assumptions, the pattern is strikingly similar to the observed phase spiral (Fig.\,\ref{fig:gaiaDR3_zvz_df}).\footnote{Note that the perturbation, $f_1 = f - f_0$, cannot be directly compared with the density contrast, $\delta f = f - \bar{f}$, used for visualizing the data, as the latter depends on the smoothing scale used to compute $\bar{f}$. In Section \ref{sec:simulation}, we will apply the same smoothing technique to the particle simulation to enable a quantitative comparison with the data.}

Our linear analysis demonstrates that a two-armed phase spiral can form even if the vertical oscillation period is much shorter than the lifetime of the galactic spiral arms, revising the conclusion of \cite{Banik2023ComprehensiveII}. This naturally explains why we observe two-armed phase spirals in the inner disk, both in the real data and in high-resolution $N$-body simulations \citep{Hunt2022Multiple,Asano2025Ripples}. Interestingly, the predicted phase spiral rotates steadily\footnote{\cite{Widrow2023Swing} studied the vertical disk dynamics within the framework of the shearing box approximation and also identified stationary two-armed phase spirals in the wake of a corotating perturber. The key difference from our model is that, in his case, the phase spiral appears stationary only when observed at a fixed position relative to the perturber: stars develop phase spirals as they stream through the wake once, but they become increasingly wound downstream of the wake. In contrast, our perturbation is periodic, and the phase spiral rotates steadily at all times at any point in the disk.}, contrary to the common notion that phase spirals are continuously winding structures.

\section{Nonlinear analysis}
\label{sec:nonlinear_theory}

\begin{figure*}
  \begin{center}
    \includegraphics[width=15.5cm]{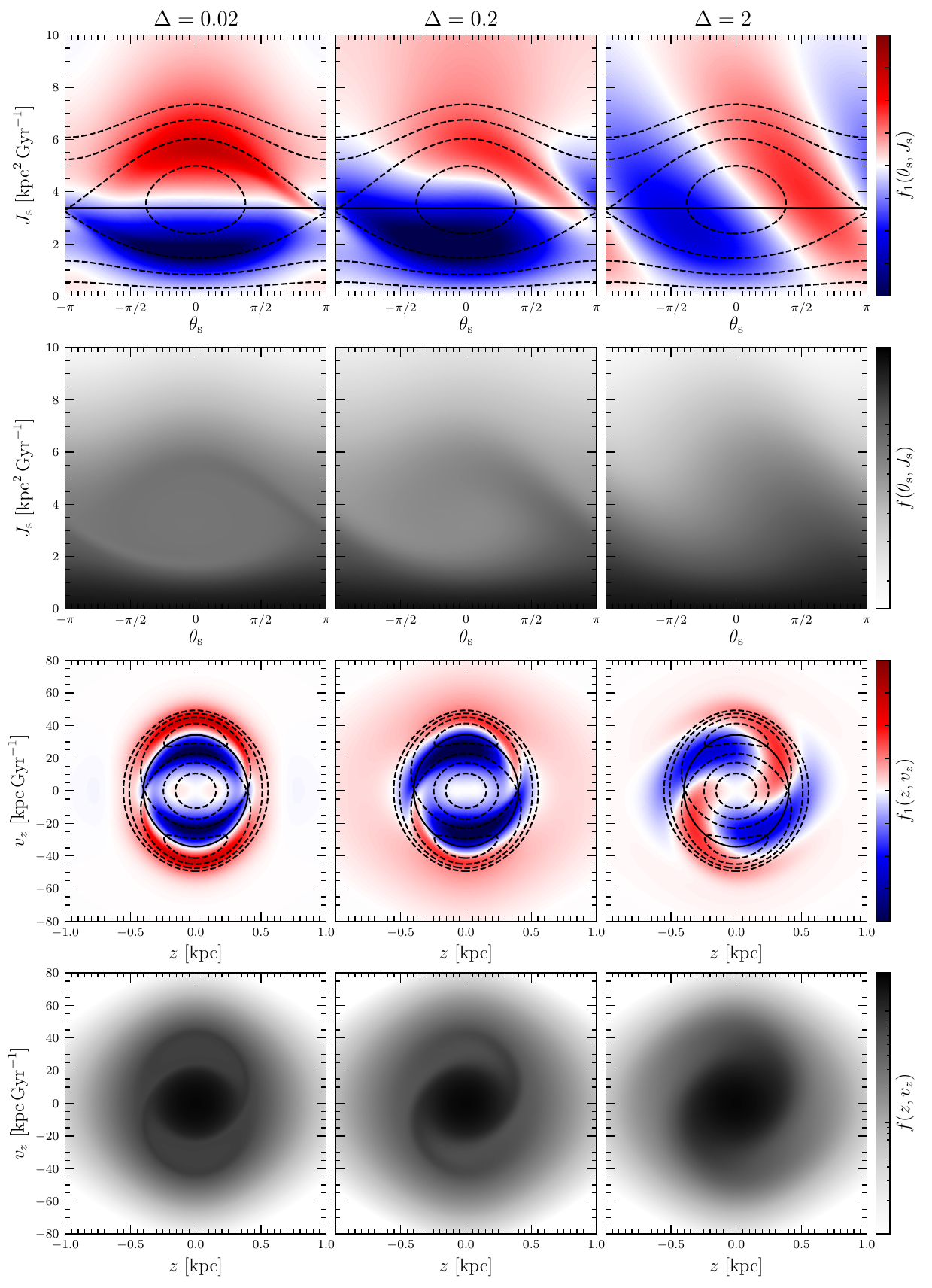}
    \caption{Steady-state solutions of the non-linearized kinetic equation (\ref{eq:kinetic_equation_slowAA}), describing the evolution of the disk subject to both persistent spiral perturbations and stochastic perturbations. We plot the perturbation $f_1$ as well as the full distribution $f=f_0+f_1$ in the slow angle-action space and in the $(z,v_z)$ space. We also overlay the contours of the Hamiltonian (\ref{eq:newHamiltonian}), which are stationary in the slow angle-action space but rotate with time in the $(z,v_z)$ space. As the strength of diffusion $\Delta$ increases from left to right, the perturbed distribution near the resonance becomes asymmetric about $\thetas = 0$, resulting in a spiral pattern in the $(z,v_z)$ space. The strength of diffusion in our standard model is $\Delta \simeq 0.2$ (middle column).}
    \label{fig:nonlinear_DF}
  \end{center}
\end{figure*}

The linear analysis in the previous section may break down near resonances if the response becomes nonlinear. As detailed in \cite{Hamilton2023BarResonanceWithDiffusion}, the validity of linear theory in the presence of diffusion is determined by the dimensionless diffusion parameter $\Delta$ (equation \ref{eq:dimensionless_diff_param}). In the collisionless limit $(\Delta=0)$, linear theory remains valid for less than a libration period before nonlinear effects, i.e., resonant trapping, become significant \citep{Chiba2022Oscillating}. In the `collisional' regime $(\Delta \gtrsim 1)$, however, the stochastic perturbations can suppress the development of nonlinear structures, keeping the system in the linear regime \citep[e.g.,][]{Pao1988Nonlinear,Catto2020Collisional,Hamilton2023BarResonanceWithDiffusion}. In this section, we examine how the disk response varies with $\Delta$ by solving the nonlinear kinetic equation. This nonlinear (non-perturbative) approach also provides key insights into the formation mechanism of the steady phase-spiral. 

Our analysis closely follows the work by \cite{Hamilton2023BarResonanceWithDiffusion}, who studied the impact of diffusion on the bar-halo resonant interaction. The basic strategy is to identify the Hamiltonian $H$ that is conserved in the absence of diffusion and then investigate numerically how the distribution of stars evolve according to $H$ when stochastic forces are added. 

We begin by Fourier expanding the perturbation (\ref{eq:spiral_potential}) in $\thetaz$ as
\begin{align}
  \Phi_1(\thetaz,\Jz,t)
  &= \mathcal{T}(t) \bigg[ \sum_{\nz} \hat{\Psi}_{\nz}(\Jz) \e^{i \nz \thetaz} \bigg] \bigg[ \sum_{l=\pm 1} \frac{1}{2}\e^{-i l \omega t} \bigg], \nonumber \\
  &= \frac{\mathcal{T}(t)}{2} \sum_{\nz,l=\pm 1} \hat{\Psi}_{\nz}(\Jz) \e^{i (\nz \thetaz - l \omega t)},
  \label{eq:spiral_potential_Fourier_expand}
\end{align}
where $\hat{\Psi}_{\nz}$ is defined in equation (\ref{eq:Zk}). Since we are interested in the steady state response against a persistent spiral arm that had emerged in the distant past, we henceforth set $\mathcal{T}(t)=1$. Near a resonance, 
\begin{align}
  \Nz \Omega_z - \omega = 0,
  \label{eq:resonance}
\end{align}
all terms in (\ref{eq:spiral_potential_Fourier_expand}) except those with $(\nz,l) = (\pm \Nz, \pm 1)$ give rise to rapid oscillations which can be averaged out. Hence, while modelling the slow nonlinear dynamics near each resonance, it suffices to consider only the corresponding resonant term
\begin{align}
  \Phi_1(\thetaz,\Jz,t)
  &= |\hat{\Psi}_{\Nz}(\Jz)| \cos\left(\Nz \thetaz - \omega t + \arg \hat{\Psi}_{\Nz} \right),
  \label{eq:spiral_potential_Fourier_expand_averaged}
\end{align}
where we have used the relation $\hat{\Psi}_{-\nz} = \hat{\Psi}_{\nz}^{\ast}$ since the function $\Psi$ is real. For the two-armed phase-spiral $\Nz=2$, the real part of $\hat{\Psi}_{2}$ is negative, while its imaginary part is zero, so $\arg \hat{\Psi}_{2} = \pi$.

We may simplify the equation by performing a canonical transformation to the now standard slow angle-action variable $(\thetas,\Js)$ \citep[e.g.,][]{LyndenBell1979BarMechanism,Tremaine1984Dynamical}:
\begin{align}
  \thetas = \Nz \thetaz - \omega t + \arg \hat{\Psi}_{\Nz}, ~~~ \Js = \Jz / \Nz
  \label{eq:slowAA}
\end{align}
using the generating function $S(\thetaz,\Js,t) = (\Nz \thetaz - \omega t + \arg \hat{\Psi}_{\Nz}) \Js$.
The new Hamiltonian is 
\begin{align}
  H'(\theta_s,\Js) &= H + \frac{\pd S}{\pd t} = H_0(\Js) - \omega \Js + |\hat{\Psi}_{\Nz}(\Js)| \cos \thetas,
  \label{eq:newHamiltonian}
\end{align}
which is time independent as we have moved to a frame rotating at the perturbing frequency $\omega$. We may thus obtain a comprehensive view of the dynamics by drawing the level curves of $H'$ in the slow angle-action space. The top row of Fig.\,\ref{fig:nonlinear_DF} illustrates these curves with dashed black. As is typical of a resonant system, there are two distinct families of orbits near the resonance: trapped and untrapped. The trapped orbits exhibit oscillatory motion in the slow angle (libration), while the untrapped orbits freely explore the full $2\pi$ range (circulation). We also plot the corresponding level curves in the $(z,v_z)$ space (Fig.\,\ref{fig:nonlinear_DF}, third row), although we caution that $H$ is not stationary here: the level curves are constantly rotating. The configuration of these curves bears a close resemblance to the contours of the Jacobi integral in barred galaxies \citep[e.g.,][]{Contopoulos1978Periodic}.

Having modeled the Hamiltonian near the resonance, we now compute the evolution of the phase-space distribution in the presence of diffusion using the kinetic equation (\ref{eq:kinetic_eq}) introduced in the previous section. Substituting the Hamiltonian (\ref{eq:newHamiltonian}), the nonlinear kinetic equation reads
\begin{align}
  \frac{\pd f_1}{\pd t} + \left(\Omegas + \frac{\pd |\hat{\Psi}_{\Nz}|}{\pd \Js} \cos \thetas \right) \frac{\pd f_1}{\pd \thetas} + |\hat{\Psi}_{\Nz}| \sin \thetas \frac{\pd (f_0+f_1)}{\pd \Js} \nonumber \\
  = C[f_0] + C[f_1],
  \label{eq:kinetic_equation_slowAA}
\end{align}
where $\Omegas = \Nz \Omegaz - \omega$. We now assume that $C[f_0]$ drives the flattening of the overall distribution on a timescale much longer than the evolution near the resonance. With this assumption, we may ignore $C[f_0]$ and seek a steady state solution of the response $f_1$ subject to the initial condition $f_1=0$ and boundary condition $f_1=0$ far from the resonance. In practice, we set the boundary at $(\Js^{\rm min},\Js^{\rm max}) = (0,30) \kpckpcGyr$ and solve the equation on a $200\times300$ grid in $(\thetas,\Js)$ space using the fourth-order Runge-Kutta method.

We note that the kinetic equation analysed in \cite{Hamilton2023BarResonanceWithDiffusion} takes a simpler form based on the pendulum approximation: the Hamiltonian is Taylor expanded around the resonance to leading order. While this approximation allows a useful analytical treatment \citep[e.g.,][]{monari2017distribution,Chiba2022Oscillating,Hamilton2023BarResonanceWithDiffusion}, its accuracy is poor when the phase flow is highly asymmetric about the resonance, as in our case, where the resonance lies close to the origin (Fig.\,\ref{fig:nonlinear_DF}). Since we are computing the kinetic equation numerically, we refrain from making the pendulum approximation and precompute the derivatives of the Hamiltonian at each grid points.

Fig.\,\ref{fig:nonlinear_DF} shows the phase-space distribution of stars at $t = 10 \Gyr$, by which the response has reached a steady state. We plot the solution for three different values of the dimensionless diffusion parameter:
\begin{align}
  \Delta \equiv \left[\frac{D\Js}{\Nz}\sqrt{-\frac{\pd \Omegas}{\pd \Js} |\hat{\Psi}_{\Nz}|^{-3}}\right]_{\Js=\Jsres},
  \label{eq:dimensionless_diff_param}
\end{align}
which is the ratio of the timescale for stars to librate around the resonance and the timescale for stars to diffuse across the resonance (see \citealt{Hamilton2023BarResonanceWithDiffusion} for details). In our standard stochastic model, $\Delta \simeq 0.2$, corresponding to the middle column of Fig.\,\ref{fig:nonlinear_DF}.

When diffusion is very weak $\Delta=0.02$ (left), libration dominates over diffusion. In this case, stars phase mix along the dashed black curves, thereby forming a nearly flat distribution within the trapped region.\footnote{In the collisionless limit $(\Delta = 0)$, this phase-mixed steady-state DF can be readily calculated by assigning to the DF the average of the unperturbed DF along the phase-space contours \citep{Binney2016Managing,binney2017orbital,monari2017distribution,Hamilton2022PhasemixedEccentricity}. The key difference in the collisional regime $(\Delta \neq 0)$ is that the DF settles into a steady state that varies along these contours.} The resulting distribution is almost symmetric about $\thetas=0$, and the corresponding pattern in the $(z,v_z)$ space appears to be a closed ring composed of two resonant islands.

At the level of diffusion expected in our Galaxy $\Delta=0.2$ (middle), phase mixing along the motion of libration gets appreciably disrupted, causing the system to settle into a steady distribution that is asymmetric about $\thetas=0$. When mapped to the $(z,v_z)$ space, this asymmetric distribution manifests as an open phase spiral, in qualitative agreement with linear theory, though quantitatively not identical, underscoring the limitation of linear theory in this regime. Physically, the periodic force by the spiral arm is constantly pushing stars at $0<\thetas<\pi$ towards large $\Js$, but these stars fail to librate back to their original position because of the stochastic forces. This results in an overdense arm (a phase spiral) emanating from the unstable fixed point at $\thetas=\pi$.

As diffusion strengthen further $\Delta=2$ (right), the phase spiral becomes weaker and less tightly wound, approaching the linear prediction, which is valid for $\Delta \gtrsim 1$ \citep{Hamilton2023BarResonanceWithDiffusion} -- strong diffusion has effectively `relinearized' the dynamics.

\section{Test-particle simulation}
\label{sec:simulation}

In this section, we study the two-armed phase spiral predicted analytically using test-particle simulations. To enable a quantitative comparison between our models and the observation, we plot the fractional density contrast $\delta f / \bar{f} = (f - \bar{f})/\bar{f}$, where $\bar{f}$ is the Gaussian-smoothed distribution obtained by the same method as in the analysis of the \textit{Gaia} data (Fig.\,\ref{fig:gaiaDR3_zvz_df}). In all simulations, we use $10^8$ particles and integrate their orbit in $(z,v_z)$ space with a time step of $0.1 \Myr$. Random kicks are applied every $10 \Myr$ (Section \ref{sec:model_stochastic_perturbation}). Here, we apply the kicks in $(z,v_z)$ space rather than in the $(q,p)$ space, although the difference is insignificant (Appendix \ref{sec:app_diffusion}).

We begin with simulations of the disk subject to a perturbation with a single frequency, as in the previous sections. We then extend our simulation to a more realistic (albeit still a 1D) model, where the perturbation has multiple frequencies, as expected along orbits with large radial motions, similar to the \textit{Gaia} sample from which the two-armed phase spirals were found (Section \ref{sec:model_periodic_perturbation}).

\subsection{Persistent spiral arm}
\label{sec:persistent_spiral_arm}

\begin{figure*}
  \begin{center}
    \includegraphics[width=17.6cm]{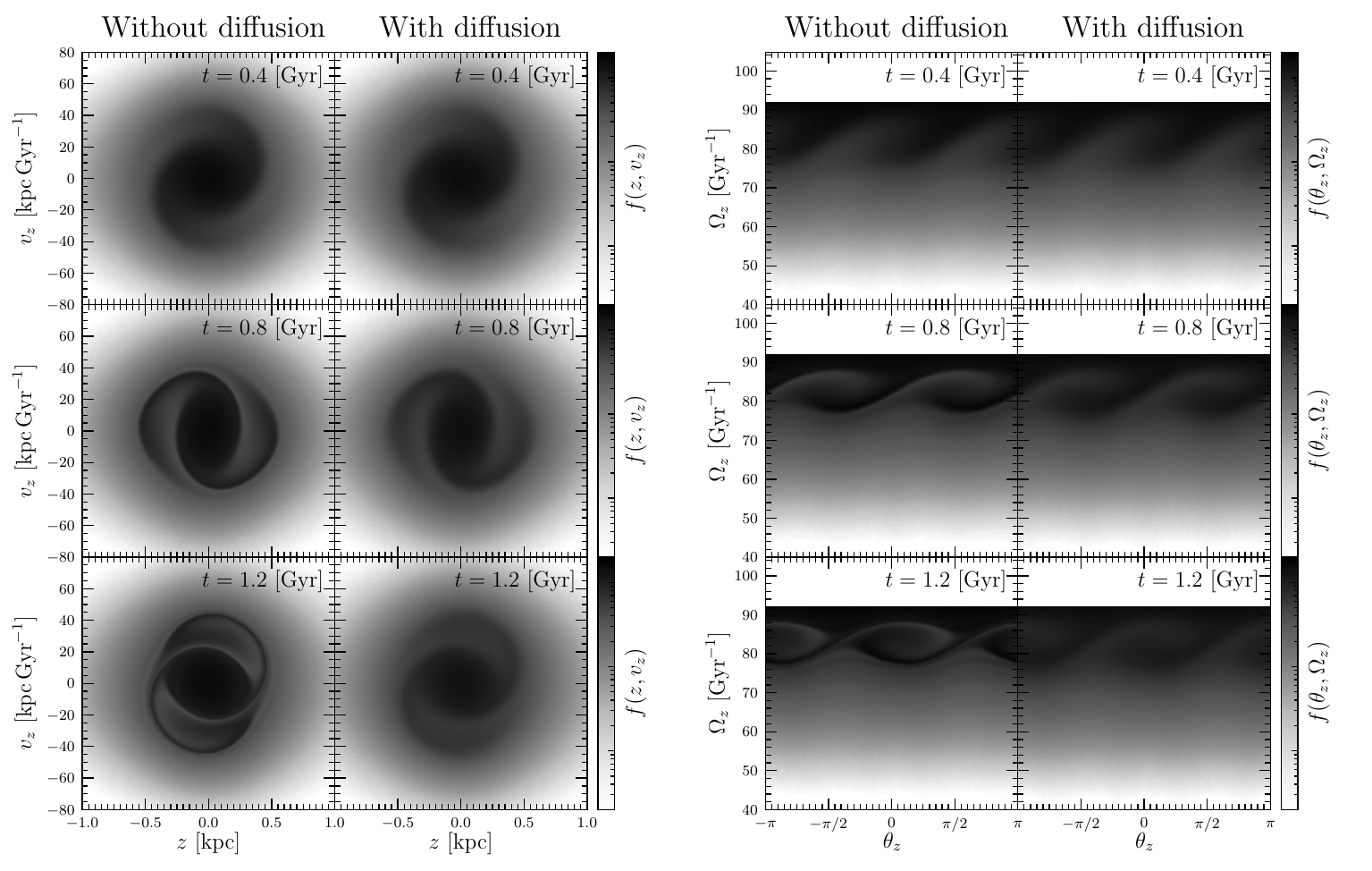}
    \includegraphics[width=17.6cm]{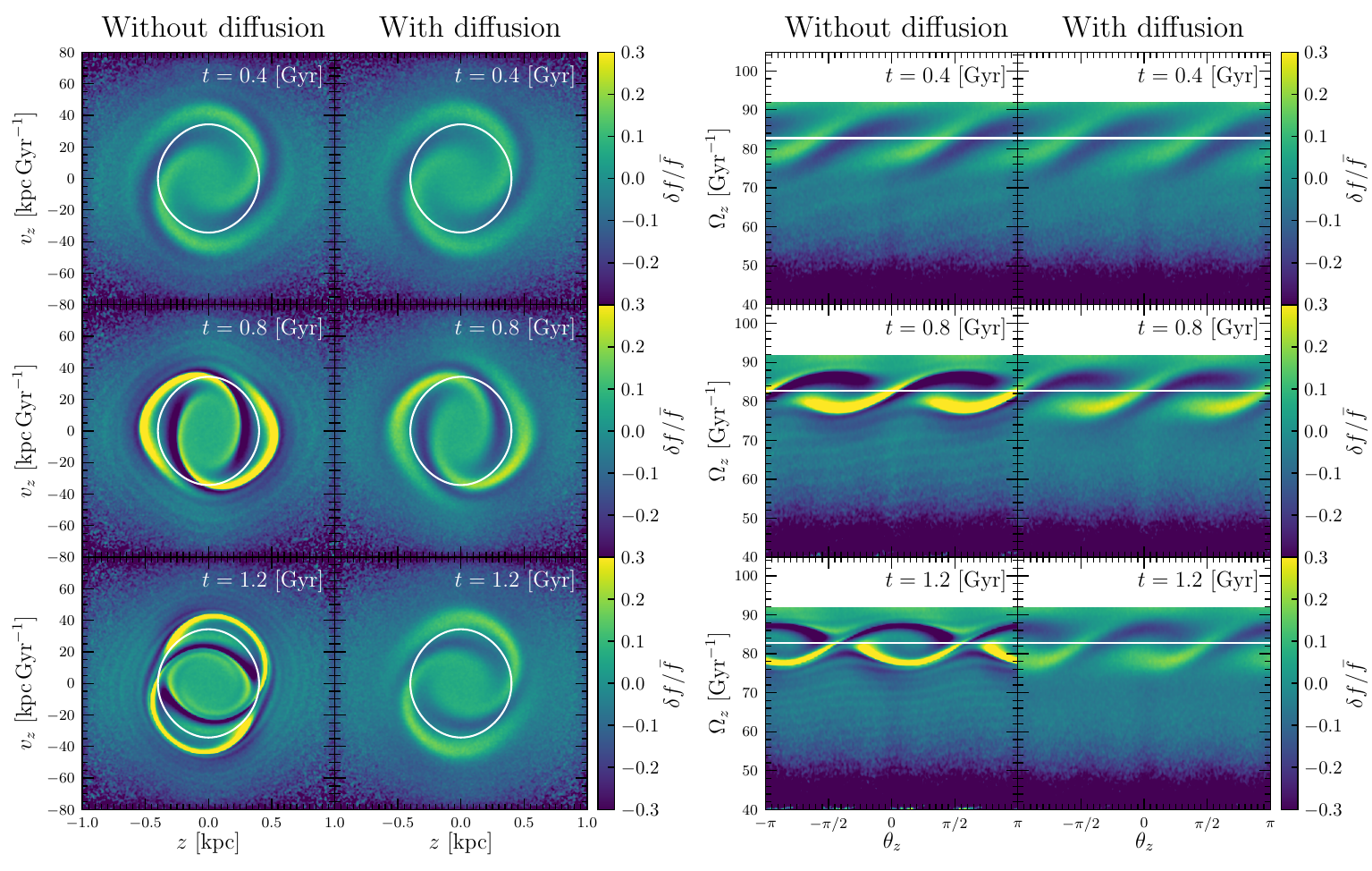}
    \caption{Test-particle simulation of the disk perturbed by a persistent spiral arm that grew with a characteristic timescale $\sigma_t = 0.2 \Gyr$ and reached a constant amplitude at $t_{\rm p}=0.4\Gyr$. The top (gray scale) panels display the full distribution $f$, while the bottom (colored) panels present the fractional density contrast relative to the smoothed distribution $\delta f/\bar{f}$, following the exact same method used to analyse the observational data (Fig.\,\ref{fig:gaiaDR3_zvz_df}). Results without and with diffusion are plotted for comparison. Diffusion disrupts the motion of libration and gives rise to a steady two-armed phase spiral near the resonance (white curves) in line with our analytical prediction.}
    \label{fig:sim_zvz_persistent}
  \end{center}
\end{figure*}

\begin{figure*}
  \begin{center}
    \includegraphics[width=17.6cm]{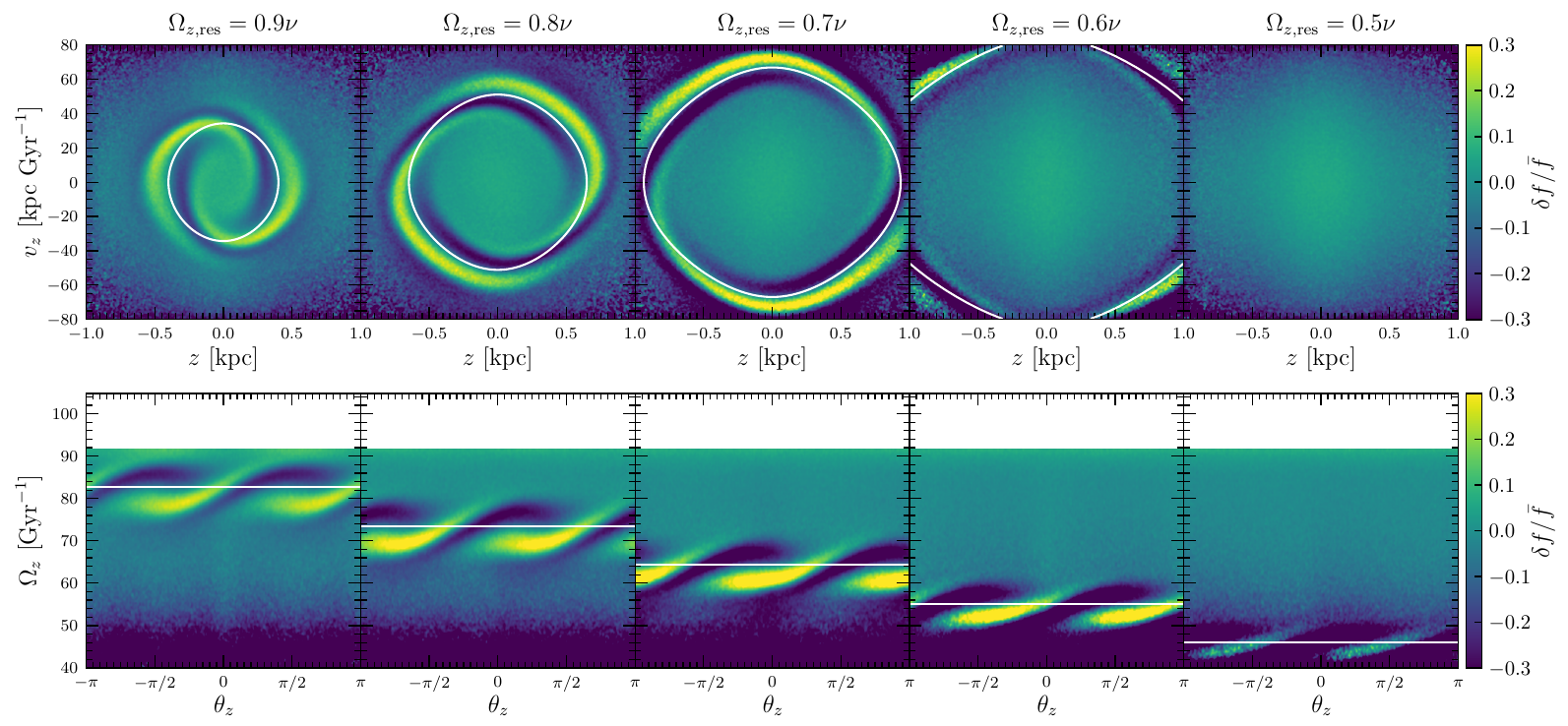}
    \caption{Disk response to persistent spiral perturbations with different perturbing frequencies $\omega=\Nz\Omega_{z,{\rm res}}$, decreasing from left to right. The snapshot is taken at $t=0.8 \Gyr$, which is $0.4 \Gyr$ after the spiral arm has fully grown. The white curves mark the location of the resonance, which shifts towards larger $\Jz$ (lower $\Omegaz$) as $\omega$ decreases. The two-armed phase spirals do not form in the absence of a resonance (rightmost plot).}
    \label{fig:sim_zvz_wnu}
  \end{center}
\end{figure*}

Fig.\,\ref{fig:sim_zvz_persistent} shows the disk response to a persistent spiral arm that grows with a characteristic timescale $\sigma_t = 0.2 \Gyr$ and reaches peak amplitude at $t_{\rm p}=0.4\Gyr$ (equation \ref{eq:spiral_potential_temporal}). The upper block (top three rows) plots the full distribution, while the lower block plots the fractional density contrast. The left two columns show the time evolution of density in the $(z,v_z)$ space, while the right show the equivalent in the $(\thetaz,\Omegaz)$ space. We compare simulations with and without diffusion as denoted on the top.

In the presence of diffusion, a two-armed phase spiral forms and rotates without winding, confirming our analytical prediction. The predicted amplitude is $\delta f / \bar{f} = 0.1-0.2$, which is consistent with or slightly larger than the observed amplitude (Fig.\,\ref{fig:gaiaDR3_zvz_df}). In the absence of diffusion, we see stars trapped in resonance slowly librating around the resonance. A phase spiral develops \textit{within} the trapped phase-space because the frequency of libration drops toward the separatrix \citep{Chiba2022Oscillating}. The distribution in the $(\thetaz,\Omegaz)$ space further clarifies the dynamics, where the trapped stars form resonant islands that steadily drift in angle. Diffusion deforms these islands into non-shearing stripes, although their shape is not perfectly straight as one would expect from a single impulsive perturbation.

Fig.\,\ref{fig:sim_zvz_wnu} plots the distribution at $t=0.8 \Gyr$ for different choices of the perturbing frequency $\omega$ in the presence of diffusion. As $\omega$ declines, the resonance shifts towards large $\Jz$ and lower $\Omegaz$, causing a corresponding shift in the phase spiral's location. The result clarifies that a two-armed phase spiral does not form in regions away from a resonance. In particular, the rightmost plot shows that, without any resonance, there would be no phase spirals.

\subsection{Transient spiral arm}
\label{sec:transient_spiral_arm}

\begin{figure*}
  \begin{center}
    \includegraphics[width=17.6cm]{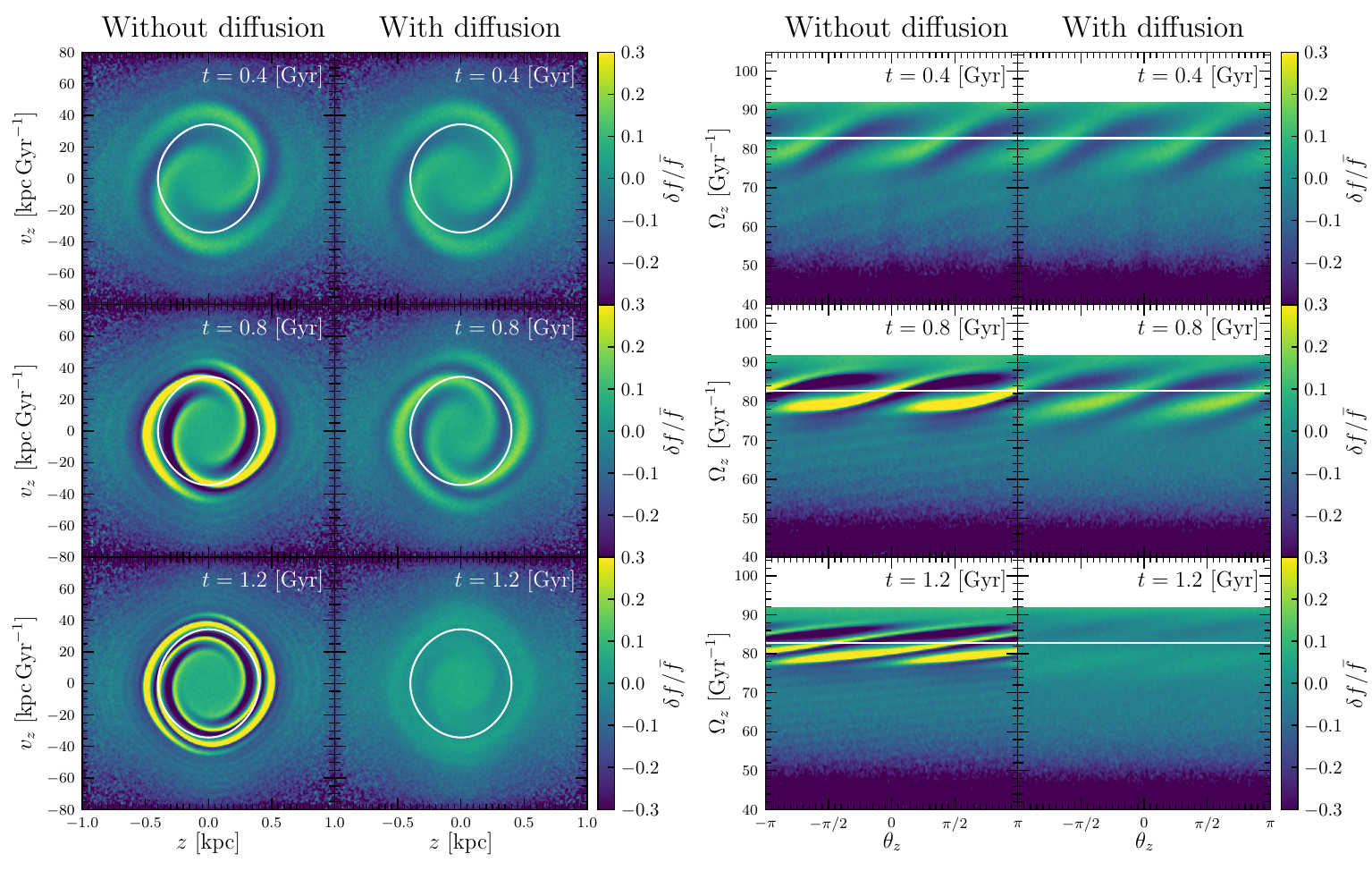}
    \caption{Disk response to a transient spiral arm that grows and decays as a Gaussian with characteristic width $\sigma_t = 0.2 \Gyr$ and peak at $t_{\rm p} = 0.4 \Gyr$. The phase spiral decays soon after the spiral arm vanishes.}
    \label{fig:sim_zvz_transient}
  \end{center}
\end{figure*}

Let us now examine what happens if the perturbation is transient, gradually decaying after reaching its peak amplitude. $N$-body simulations suggest that spiral arms are indeed transient features that recurrently grow and decay with lifetimes typically ranging from a few hundred million to a few billion years \citep[e.g.,][]{Sellwood1984Spiral,SellwoodCarlberg2014,Baba2009Origin,Sellwood2011lifetimes,Fujii2011Dynamics,Grand2012Dynamics,Roskar2012Radial,DOnghia2013ASelfperpetuating}. 

Fig.\,\ref{fig:sim_zvz_transient} shows an example of the disk response to a transient spiral arm that varies as a Gaussian with width $\sigma_t = 0.2 \Gyr$ and peak at $t_{\rm p} = 0.4 \Gyr$. A phase spiral appears temporarily just after the spiral arm fully grows. As the spiral perturbation decays, stars are released from the resonance, and consequently the phase spiral winds up. In the absence of diffusion, the phase spiral continues to wind indefinitely, whereas with diffusion, the phase spiral vanishes soon after the perturbation disappears. This behavior is consistent with the $N$-body simulations by \cite{Hunt2022Multiple}, who report that a series of two armed phase-spirals appear and disappear across the disk as spiral structures form and evolve.

\begin{figure*}
  \begin{center}
    \includegraphics[width=17.6cm]{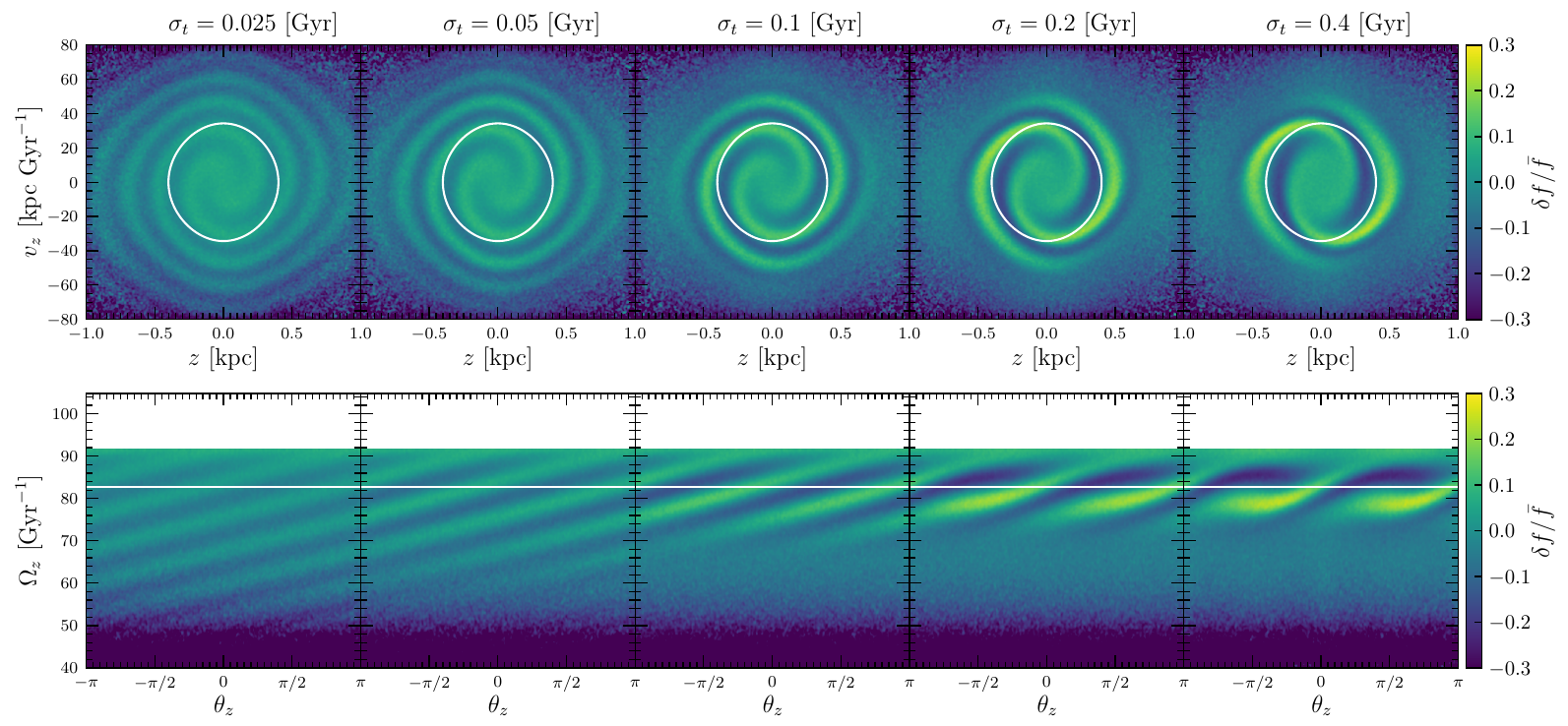}
    \caption{Dependence of disk response on the characteristic lifetime of the perturbation $\sigma_t$, increasing from left to right. The perturbation peaks at $t_{\rm p}=0.4 \Gyr$ and the snapshot is taken at $t=0.8 \Gyr$. As the lifetime shortens, the two-armed phase spiral weakens and the region over which it forms expands, in agreement with theory \citep{Banik2023ComprehensiveII}. As the lifetime increases, the response becomes nonlinear, approaching the response to a persistent perturbation (Fig.\,\ref{fig:sim_zvz_persistent}).}
    \label{fig:sim_zvz_sigt}
  \end{center}
\end{figure*}

Fig.\,\ref{fig:sim_zvz_sigt} plots the disk response at $t=0.8 \Gyr$ for different choices of the spiral's characteristic lifetime $\sigma_t$. As $\sigma_t$ decreases, the range of frequency over which the phase spirals appears broadens, since the perturbation becomes less adiabatic away from the resonance. The response also weakens with decreasing $\sigma_t$ because, in the impulsive regime ($\sigma_t \rightarrow 0$), the total work done on the system scales with the duration of the applied force (\citealt{Banik2023ComprehensiveII}).

\subsection{Realistic model with multiple resonances}
\label{sec:simulation_Jr80}

So far we have employed a periodic perturbation with a single perturbing frequency. We now extend our model to a more realistic perturbation with multiple perturbing frequencies. This is motivated by the fact that the two-armed phase spiral is found in the phase-space distribution of Solar neighborhood stars with relatively small angular momentum and large radial action (Section \ref{sec:model_periodic_perturbation}): along such an eccentric orbit, the temporal spectra of the spiral arms' potential contains multiple peaks at high-order resonances (Fig.\,\ref{fig:spectra}). As detailed in Section \ref{sec:model_periodic_perturbation}, we keep our simulation one dimensional in the vertical direction, but we determine the frequencies of the potential perturbation by integrating a single orbit in a two-dimensional Mestel disk and recording the time variation of the spiral arm's potential along that orbit. As in Section \ref{sec:model_periodic_perturbation}, we adopt $\Sigma_{\rm max}=5.5 \Msun\pc^{-2}$, $m=2$, $\alpha=12^\circ$, $\beta=0.5$, $\Omegap=28 \Gyr^{-1}$ by default.

\begin{figure*}
  \begin{center}
    \includegraphics[width=17.6cm]{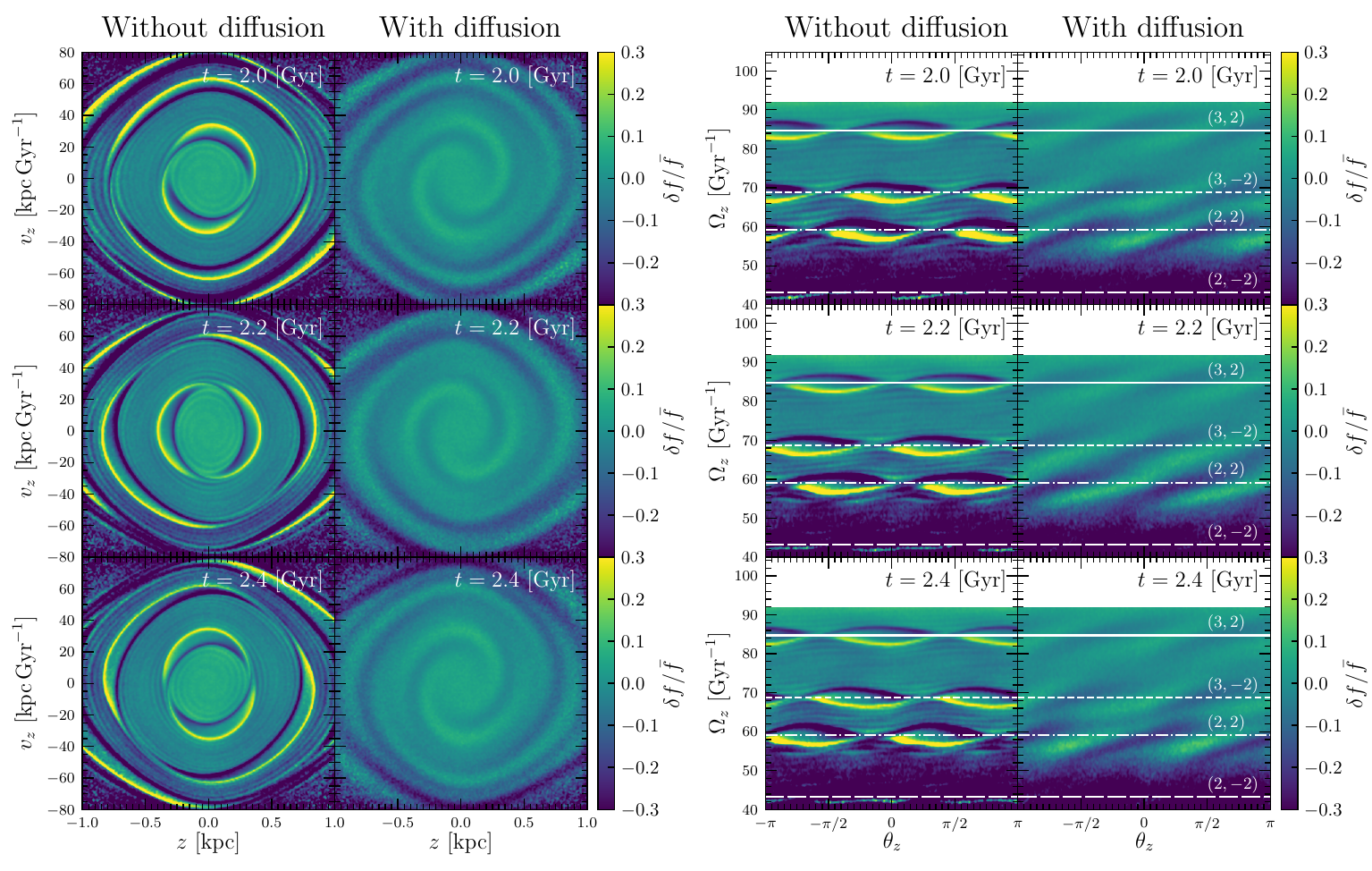}
    \caption{Vertical phase-space distribution of stars on eccentric orbits ($\JR = 80 \kpckpcGyr$) with relatively small angular momentum ($\Jphi = 1488 \kpckpcGyr$), similar to the \textit{Gaia} samples from which the two-armed phase-spiral was found. The star's large radial motion results in  multiple resonances. The white lines mark the location of the $\Nz=2$ resonances, where the resonant indices $(\NR,\Nphi)$ are denoted in the rightmost column. We did not mark the resonances in the $(z,v_z)$ space as they hinder the perception of the phase spirals.}
    \label{fig:sim_zvz_Jr80}
  \end{center}
\end{figure*}

Fig.\,\ref{fig:sim_zvz_Jr80} shows the response of stars on orbits with relatively small angular momentum, $\Jphi = 1488 \kpckpcGyr$, and large radial action, $\JR=80\kpckpcGyr$, mimicking the \textit{Gaia} sample. As in Section \ref{sec:persistent_spiral_arm}, the spiral arm is persistent, growing with a characteristic timescale $\sigma_t = 0.2 \Gyr$ until $t_{\rm p}=0.4\Gyr$. As clear from the non-diffusive case, there are multiple resonances each creating their own chain of resonant islands, where the resonant indices are indicated on the rightmost column. These resonant islands are smaller than those in the single-frequency model (e.g., Fig.~\ref{fig:sim_zvz_persistent}), as the power of the perturbation is spread out across multiple resonances. When diffusion is present, the resonant islands become smeared, transforming into multiple local phase-spirals. Since each local phase-spiral rotates at distinct resonant frequencies, they connect, disconnect, and reconnect with each other over time, forming a global phase-spiral that temporarily appears to wind up. This phenomenon is reminiscent of the evolution of spiral arms themselves, which are composed of several modes with unique pattern speeds that together give rise to an apparently shearing pattern \citep{SellwoodCarlberg2014,Sellwood2019SpiralInstabilities,Sellwood2021Spiral}.

\begin{figure}
  \begin{center}
    \includegraphics[width=8.5cm]{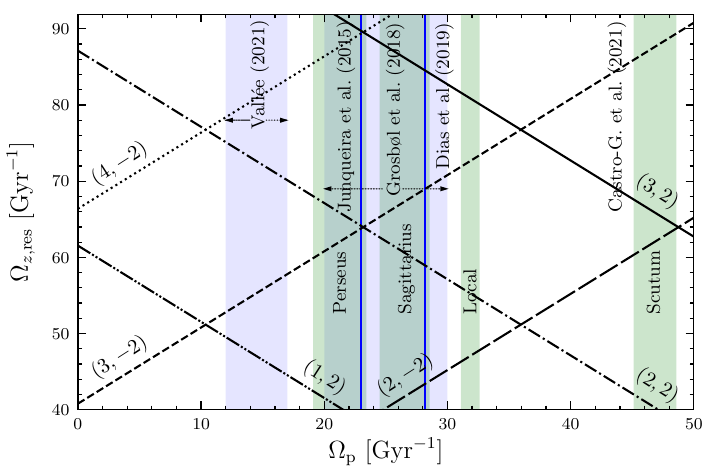}
    \caption{Possible resonances that could generate a two-armed phase spiral in the inner Galactic disk ($\Rg=6.2\kpc$). The black inclined lines mark the location of the $\Nz=2$ resonances, where the brackets denote $(\NR,\Nphi)$. The vertical lines and bands indicate the measured pattern speeds of spiral arms in the Milky Way.}
    \label{fig:best_omegap}
  \end{center}
\end{figure}

The shape of the phase spirals is dependent on the assumed pattern speed, which sets the location of the resonances. Fig.\,\ref{fig:best_omegap} illustrates how the location of the resonances varies as a function of the pattern speed, where the brackets denote $(\NR,\Nphi)$. Resonances with positive $\Nphi$ shift downward with increasing $\Omegap$, while the opposite occurs for resonances with negative $\Nphi$. The vertical lines and bands mark the measured spiral pattern speeds in the Milky Way, where blue indicates measurements of a single global pattern speed \citep{Junqueira2015new,Grosbol2018spiral,Dias2019spiral,Vallee2021low}, while green indicates recent measurements of individual spiral arms \citep{CastroGinard2021Milky}.\footnote{Some studies find little difference in the pattern speeds of individual spiral arms \citep[e.g.,][]{Dias2019spiral,Monteiro2021distribution}.} With low values of pattern speeds, $10 \lesssim \Omegap \lesssim 23 \Gyr^{-1}$, the resonances that drive the phase spirals are $(\NR,\Nphi)=(4,-2),(2,2),(3,-2),(1,2)$, from top to bottom. With intermediate pattern speeds, $23 \lesssim \Omegap \lesssim 36 \Gyr^{-1}$, the drivers of the phase spirals shift to $(\NR,\Nphi)=(3,\pm2),(2,\pm2)$. With high pattern speeds, $\Omegap \gtrsim 36\Gyr^{-1}$, the driving resonances are again $(\NR,\Nphi)=(3,\pm2),(2,\pm2)$ but the order of each set of resonances are reversed.

\begin{figure*}
  \begin{center}
    \includegraphics[width=17.6cm]{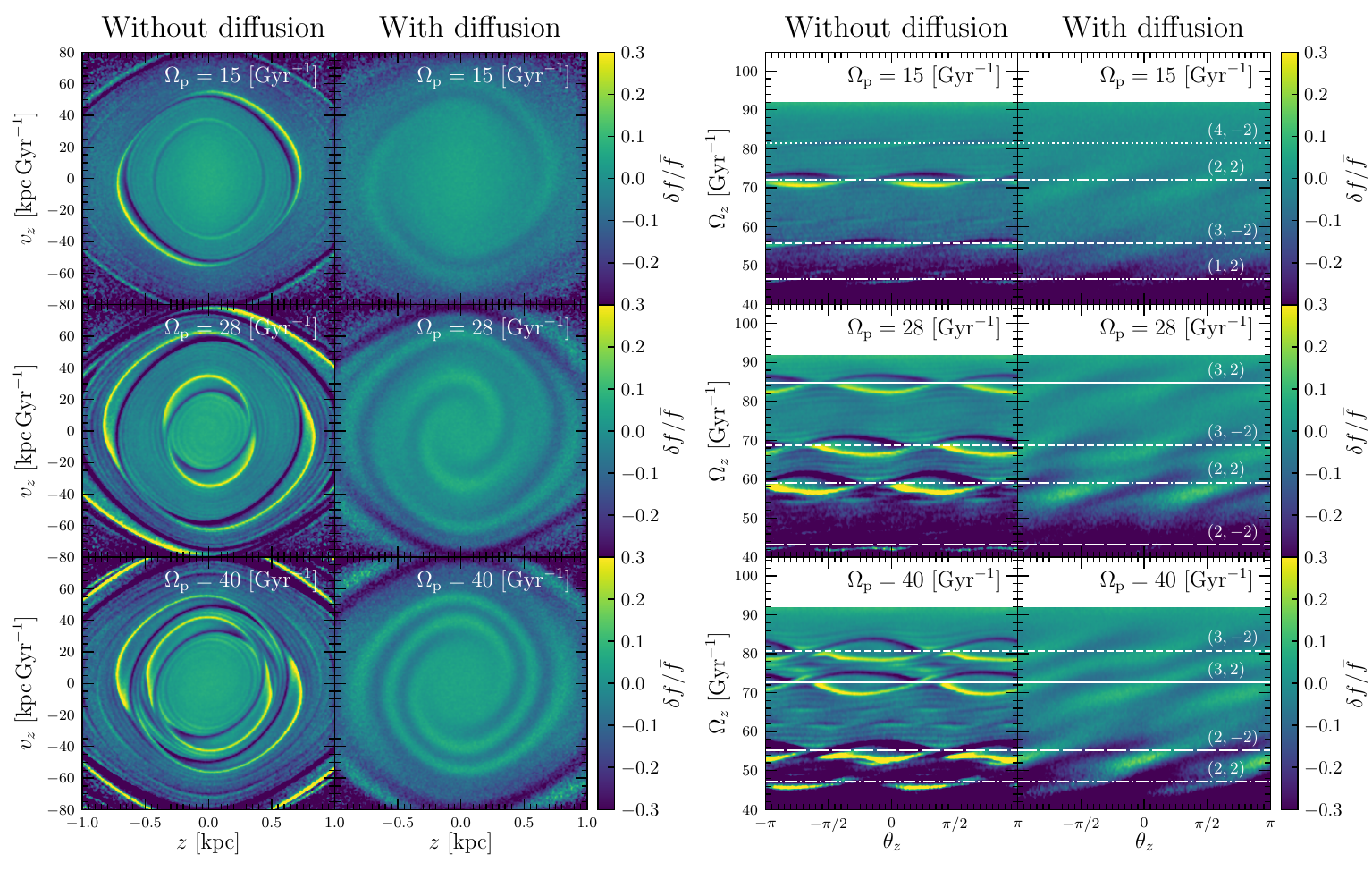}
    \caption{Similar to Fig.\,\ref{fig:sim_zvz_Jr80}, but with various pattern speeds of spiral arms. The snapshot is taken at $t=2.4 \Gyr$. The white lines mark the location of the $\Nz=2$ resonances, where the resonant indices $(\NR,\Nphi)$ are denoted in the rightmost column.}
    \label{fig:sim_zvz_Jr80_wp}
  \end{center}
\end{figure*}

Fig.\,\ref{fig:sim_zvz_Jr80_wp} shows the response for three different spiral pattern speeds: $\Omegap=15$, $28$, and $40 \Gyr^{-1}$. The snapshot is taken at $t=2.4 \Gyr$. As discussed, the resonances shift in frequency and rearrange their order as the pattern speed changes. When two resonances are closely spaced (bottom row), they can form a smoothly connected phase spiral, which could be mistakenly interpreted as the result of an impulsive perturbation (cf. Fig.~\ref{fig:sim_zvz_sigt}, leftmost plot).

The strength of the perturbation also varies with the pattern speed, since the spiral's amplitude decays with distance from the corotation radius (equation \ref{eq:spiral_potential_2d}). For example, with $\Omegap=15 \Gyr^{-1}$, the corotation radius lies far out at $\RCR = 16 \kpc$, so the spiral amplitude at $R = 6.2 \kpc$ is $22\%$ of the maximum value at corotation, assuming $\beta=0.5$. This corresponds to a $47\%$ reduction in the width of the resonant islands, which scales as the square root of the perturbation \citep[e.g.,][]{lichtenberg1992regular}. With $\Omegap=28 \Gyr^{-1}$ ($\RCR = 8.6 \kpc$), the perturbation amplitude (resonant width) at $R = 6.2 \kpc$ is reduced by $74 \%$ ($86\%$) from the peak value.

Our model suggests that galactic spiral arms with a range of pattern speeds can generate two-armed phase spirals in the inner disk. Spiral arms with a low pattern speed $\Omegap \sim 15 \Gyr^{-1}$ \citep[e.g.,][]{Vallee2021low,Khalil2025nonaxisymmetric}, which place corotation far out at $\RCR \sim 16 \kpc$, are less favored, since their influence in the inner disk would be weak if, as we assume, their amplitude declines with distance from corotation. In our current framework, the spiral arm most likely responsible for the observed two-armed phase spiral is the Sagittarius-Carina arm which has pattern speed $\Omegap = 26-30 \Gyr^{-1}$ and lies at $R = 6-8 \kpc$ \citep{CastroGinard2021Milky,Monteiro2021distribution,Joshi2023Revisiting}. The presence of other spiral arms would make the phase space packed even more densely with resonances. Fitting the shape and amplitude of the phase spiral could place independent constraints on the spirals' pattern speeds, though this is beyond the scope of this paper.

\section{Discussion}
\label{sec:discussion}

\subsection{Phase-space structure preserved by diffusion}
\label{sec:diffusion_preserve}

An important physical insight from our work is the crucial role played by diffusion in \textit{preserving} (rather than destroying) phase spirals. Persistent periodic perturbations initially drive phase spirals near resonances, but it is diffusion that maintains them over time by preventing trapped stars from phase mixing and making the distribution symmetric in the slow angle \citep[Section \ref{sec:nonlinear_theory}, see also][]{Hamilton2023BarResonanceWithDiffusion}. This stands in contrast to the earlier work by \cite{Tremaine2023Snail}, where diffusion acted to wipe out the phase spirals. The key difference is that they considered stochastic white-noise forcing — a series of uncorrelated large-scale impulsive kicks — rather than persistent periodic forcing. In their model, the phase spirals were generated by the impulsive kicks, and the role of diffusion was merely to destroy them. Our result suggests that, contrary to earlier expectations, diffusion may not always act to erase structures in phase space; instead, under certain conditions, diffusion can actively preserve phase-space structures, with implications beyond the vertical phase-spiral.

\subsection{Two-armed phase spirals in the Milky Way}
\label{sec:two_armed_phase_spirals_Milky_Way}

Since spiral arms are prevalent throughout the galactic disk and there are many resonances with similar strength (Fig.\,\ref{fig:spectra}), our results suggest that the two-armed phase spiral should be excited ubiquitously across the disk. Indeed, \cite{Hunt2022Multiple} found in their high-resolution simulation of an isolated galaxy that two-armed phase spirals emerge over a range of radii as the disk undergoes secular evolution. This raises the question of why the two-armed phase spiral has been observed only in the inner disk and not in the Solar neighborhood and beyond. \textit{Gaia}'s observation suggests that the overall amplitude of the combined spiral structure does not decline strongly with radius, even though individual arms may peak near their respective corotation radii \citep{Eilers2020Spiral,GaiaDR32023Mapping}. Moreover, it is unlikely that the two-armed phase spiral is obscured by the more dominant one-armed phase spiral in the outer disk: \cite{Frankel2023Vertical} found no clear evidence of a two-armed phase spiral in the residual between the data and the one-armed model. Furthermore, stellar diffusion is expected to be stronger in the inner disk, where the density of giant molecular clouds is higher. It is therefore nontrivial that we observe the two-armed phase spirals only in the inner disk. This remains a topic for future investigation.

\subsection{Galactic bar}
\label{sec:galactic_bar}

Similar to spiral arms, the bar in our Galaxy may also generate two-armed phase spirals through its resonances. However, within the framework of our simple model, the bar is disfavored as the main driver of the observed pattern, because there are no strong vertical resonances near the orbits of stars forming the two-armed phase spiral: The guiding radius of these stars is $\Rg \simeq 6.2\kpc$, which lies close to the bar's corotation radius, $\RCR = 5.8-6.9 \kpc$ \citep[e.g.,][]{Binney2020Trapped,Chiba2021TreeRing,Clarke2021ViracGaia,Lucey2023Constraining,Leung2023measurement,Zhang2024Kinematics,Dillamore2025Dynamical}, meaning that the azimuthal frequency of these stars with respect to the bar ($\Omegaphi-\Omegap$) is very low. Hence, to have a vertical resonance at a high vertical frequency, the vertical motion of these stars must resonate with their radial motion. However, unlike spiral arms, the bar's vertical potential does not have a strong radial dependence, so the resonances involving radial motions ($\NR \neq 0$) are much weaker than that of spiral arms (Appendix \ref{sec:app_bar_resonance}). This renders the bar difficult to create a strong phase spiral at the observed phase space.

The consideration above relies on a variety of assumptions made in our model. Of these, we have assumed that the perturber has a fixed pattern speed. Using test-particle simulations, \cite{Li2023Gaia} demonstrates that a bar can temporarily generate a pronounced two-armed phase spiral if its pattern speed decreases over time, causing the resonances to constantly move. A moving resonance indeed leaves behind a striated pattern in the slow angle-action space \citep{Chiba2023GeneralFastSlowRegime}, which may appear as a spiral in the $(z,v_z)$ space. It would be interesting to see whether and for how long these patterns survive in a realistic, noisy environment.

\subsection{Limitations of current model}
\label{sec:limitations}

This work employed the simplest possible model to describe the physics behind the formation of the two-armed vertical phase-spiral. Given its simplicity, its scope is necessarily limited. Below, we outline some of the aspects that warrant investigation in future work.

First, we have modelled the vertical structure of the Galactic disk with an isothermal slab, in which the DF is an exponential function of the vertical energy $E_z$. While the DF of the thick disk is nearly isothermal \citep[e.g.,][]{Cheng2024surface}, that of the thin disk is known to decay more rapidly with $E_z$. For example, \cite{Binney2023Selfconsistent} find a good fit to the data with a DF exponential in the vertical action $J_z$. This implies that we may have underestimated the amplitude of the phase spiral, since the resonant response directly scales with the gradient of the unperturbed DF with respect to $J_z$ (e.g., equation \ref{eq:kinetic_eq_lin_sol}). There is also a non-negligible difference between the potential of the isothermal model and that of the Milky Way (Fig.\,\ref{fig:Phi}), which could cause small shifts in the resonance locations. Although neither effect would qualitatively alter our conclusions, they could become important when performing quantitative comparisons with the data.

Second, we have restricted our model to one dimension perpendicular to the Galactic plane and focused on the vertical perturbation by the Galactic spiral arms. In reality, however, spiral arms also induce significant horizontal (in-plane) perturbations, changing the stars' angular momentum $\Jphi$. Because the star's vertical oscillation frequency $\Omegaz$ depends strongly on $\Jphi$, these horizontal perturbations can indirectly modulate the vertical dynamics. In addition, stars phase mix not only in the vertical direction due to variations in $\Omegaz$, but also horizontally due to variations in $\Omegaphi$ and $\Omegar$. When the DF is course-grained over a finite range in angular momentum, this horizontal mixing results in a $1/t$ decay of the response \citep{Banik2023ComprehensiveII}. Since this decay is slower than the super-exponential due to small-scale diffusion, horizontal mixing effectively imposes a `floor' to the diffusion rate--—one that would persist even in the absence of molecular clouds.

Finally, we have ignored collective effects, i.e., we ignored the gravitational potential generated self-consistently by the perturbed part of the stellar distribution. Previous studies on linear, collisionless disks have shown that collective effects give rise to sustained breathing modes \citep{Weinberg1991Vertical} and can generate strong two-armed vertical phase-spirals through swing amplification \citep{Widrow2023Swing}. How collective effects operate in nonlinear, collisional systems such as ours remains unclear and would be a topic for future work.

\section{Summary}
\label{sec:summary}

We studied the mechanism by which galactic spiral arms induce two-armed phase spirals in the vertical motion of the stellar disk. While past theories predict that the formation of such a phase spiral requires a non-adiabatic (impulsive) perturbation that grows and decays over a timescale much shorter than the vertical oscillation period ($\lesssim 0.1 \Gyr$), we show that spiral arms with a realistically long lifetime ($\gtrsim 0.1 \Gyr$) can in fact give rise to a two-armed phase spiral if there are (i) resonances, which break the adiabaticity, and (ii) stochastic kicks (e.g., due to molecular cloud scattering), which break the angular symmetry of the perturbed distribution, turning ring-like resonant structures into open spirals.

The formation process of the two-armed phase spiral is best understood by first considering an ideal case where the disk is persistently perturbed by a potential that is symmetric in $z$. In this case, stars with vertical oscillation frequencies close to the perturbing frequency become resonantly trapped and librate around the resonance. Over time, the distribution of these trapped stars phase-mixes along the perturbed orbit, forming a new equilibrium distribution with a ring-like structure that rotates steadily in the $(z,v_z)$ space at the resonant frequency. When a stochastic force is introduced, these stars diffuse through phase space, preventing them from completing a full cycle of libration. As a result, phase mixing remains incomplete. This leads to a steady distribution that is asymmetric in phase with respect to the center of libration ($\thetas=0$): due to the negative gradient in the initial distribution function, the density of stars pushed towards large action at $\thetas>0$ is always higher than that of those pulled towards small action at $\thetas<0$. When this asymmetric distribution is projected onto the $(z,v_z)$ space, it manifests as an open phase spiral that rotates steadily without winding.

For transient spiral arms, the phase spiral induced near the resonance winds up and vanishes once the spiral arms disappear. In this situation, the role of diffusion in forming the phase spiral becomes minor. However, resonance still remains essential: for the phase spiral to form far from the resonance, the characteristic lifetime of the perturbation must be unrealistically short (Fig.\,\ref{fig:sim_zvz_sigt}). Furthermore, the amplitude of the phase spiral generated by such an extremely short-lived spiral arm is too weak, since the total work done by the perturbation scales with its lifetime.

In realistic systems, we expect the vertical phase-space to host a high density of closely spaced resonances because there are possibly multiple spiral arms with different pattern speeds, and also because the temporal spectra of the spiral perturbation has large powers at high-order resonances when the orbit has a large radial motion (Fig.\,\ref{fig:spectra}). Since each resonance creates a local phase spiral, which rotates steadily at distinct frequencies, they give rise to a global spiral pattern in the $(z,v_z)$ space that appears to be shearing, in close analogy with the dynamics of the spiral arms themselves \citep{Sellwood2019SpiralInstabilities}.

We find that a spiral arm with surface density amplitude recently measured by \cite{Eilers2020Spiral} and pattern speed $\Omegap=28 \Gyr^{-1}$, corresponding to the Sagittarius-Carina arm, can naturally generate two-armed phase spirals with amplitudes comparable to those observed. More detailed modelling of the data would require careful treatment of selection effects \citep{Frankel2023Vertical}, which can introduce $\Nz=2$ signals, as well as accounting for the amplification of the response by self-gravity \citep{Widrow2023Swing} and the damping effects of horizontal mixing \citep{Banik2022ComprehensiveI,Banik2023ComprehensiveII}.

Our study highlights the crucial role of resonances in the formation of two-armed phase spirals. Although phase spirals in real noisy environment may not show obvious signs of resonances, removing the effects of stellar diffusion reveals that they are in fact structured by discrete resonances. Uncovering these hidden phase-space structures can shed light into the nature of both the periodic perturbation and the stochastic perturbation.

\section*{Acknowledgements}

We thank Scott Tremaine for helpful comments and encouragements. We are also grateful to Uddipan Banik for many stimulating discussions. R.C. is also thankful to the members of CLAP and MSSL for fruitful discussions and hospitality during his visit. R.C. and N.F. are supported by the Natural Sciences and Engineering Research Council of Canada (NSERC), [funding reference \#DIS-2022-568580]. R.C. also acknowledges the financial support by the Japan Society for the Promotion of Science (JSPS) Research Fellowship, grant No. 25KJ0049. C.H. is supported by the John N. Bahcall Fellowship Fund and the Sivian Fund at the Institute for Advanced Study.

Computations were performed on the Niagara supercomputer at the SciNet HPC Consortium. SciNet is funded by Innovation, Science and Economic Development Canada; the Digital Research Alliance of Canada; the Ontario Research Fund: Research Excellence; and the University of Toronto.

\section*{Data availability}

The codes used to produce the results are available from the corresponding author upon request. 



\bibliographystyle{mnras}
\bibliography{references}



\appendix

\section{Resonances of galactic bar}
\label{sec:app_bar_resonance}

Our study demonstrates that two-armed phase spirals can be generated by the combined effect of periodic resonant perturbations and stochastic perturbations. We attributed the former to galactic spiral arms, though, in principle, the galactic bar could play a similar role if it can strongly resonate with the vertical motion of stars in the disk. Here, we investigate whether such a strong resonance exists.

\begin{figure*}
  \begin{center}
    \includegraphics[width=17cm]{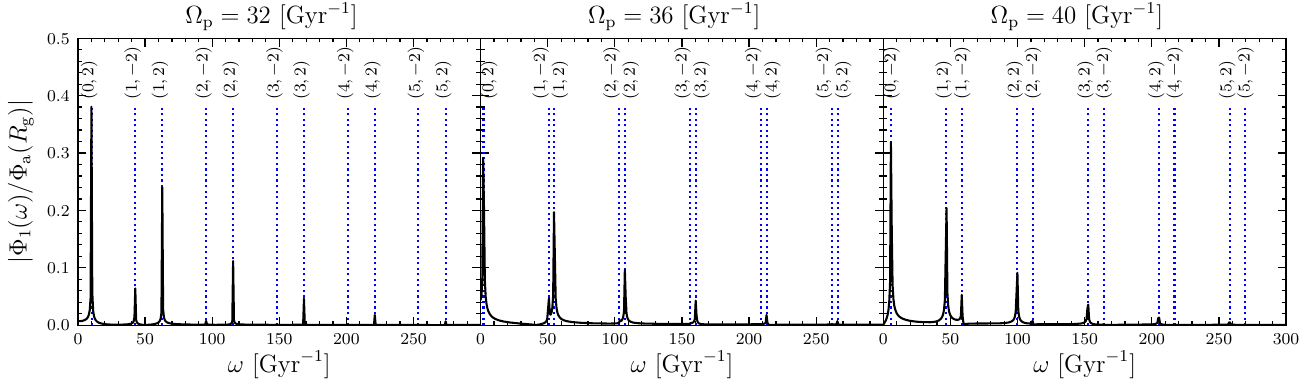}
    \caption{Spectra of the bar's potential along an in-plane unperturbed orbit with $(\JR,\Jphi,\Jz)=(80,1440,0)\kpckpcGyr$. The actions $(\JR,\Jphi)$ correspond to those of the Solar neighborhood stars from which the two-armed phase spiral was detected. The spectra are shown for three different bar pattern speeds. The blue dotted lines mark the frequencies $\omega=\NR\OmegaR+\Nphi(\Omegaphi-\Omegap)$, where the indices $(\NR,\Nphi)$ are denoted in the figure. Because the bar has a weak radial dependence, the amplitude of the high-order resonances involving the radial motion (i.e., $\NR \neq 0$) is small compared to that of the spiral arm (Fig.\,\ref{fig:spectra}, right plot).}
    \label{fig:spectra_bar_wp}
  \end{center}
\end{figure*}

Fig.\,\ref{fig:spectra_bar_wp} shows the spectra of the bar's potential presented in \cite{Chiba2020ResonanceSweeping} along an in-plane unperturbed orbit with $(\JR,\Jphi,\Jz)=(80,1440,0)\kpckpcGyr$, which is the action of stars from which the two-armed phase spiral was found. We show results for three different bar pattern speeds favoured by recent measurements \citep[e.g.,][]{Binney2020Trapped,Chiba2021TreeRing,Clarke2021ViracGaia,Lucey2023Constraining,Leung2023measurement,Zhang2024Kinematics,Dillamore2025Dynamical}. As in Fig.\,\ref{fig:spectra}, the brackets denote the set of integers $(\NR,\Nphi)$ corresponding to the frequency $\omega=\NR\OmegaR+\Nphi(\Omegaphi-\Omegap)$. The spectra are significantly biased toward small $\omega$, indicating that strong resonant perturbations are only expected at low vertical frequency ($\Omegazres=\omega/\Nz$), i.e., far from the origin of the $(z,v_z)$ space. The strongest resonances $(\NR,\Nphi)=(0,\pm2)$, directly coupling the vertical motion with the azimuthal motion, occur at very small $\omega$ $(< 10 \Gyr^{-1})$ because the guiding radius of the orbit $(\Rg = 6.2 \kpc)$ is close to the bar's corotation radius $(\RCR = 5.8-6.9 \kpc)$, i.e., $\Omegaphi-\Omegap$ is small. The resonances involving the radial motion of stars $(\NR \neq 0)$ occur at large $\omega$, but they are significantly weak because the bar's potential varies relatively little with radius as compared to spiral arms, which are tightly wound (Fig.\,\ref{fig:spectra}). We thus disfavor the bar as the main driver of the observed two-armed phase spiral. Future studies with a realistic 3D barred galaxy model \citep[e.g.,][]{Dehnen2023potentialdensitybar} will be required to confirm this conclusion.

\section{Model of spiral arm}
\label{sec:app_spiral_model}

In this appendix, we describe our model for the spiral arm. We consider the following potential-density pair:
\begin{align}
  \Phi_1(R,\varphi,z,t) &= \xi(z) \Phi_{\rm a}(R) \cos \left[ m (\varphi - \Omegap t + \cot \alpha \ln R) \right], \\
  \rho_1(R,\varphi,z,t) &= \zeta(z) \Sigma_{\rm a}(R) \cos \left[ m (\varphi - \Omegap t + \cot \alpha \ln R) \right],
  \label{eq:spiral_density}
\end{align}
where $\zeta(z)$ describes the vertical density profile of the spiral, normalized such that $\int_{-\infty}^{\infty} \drm z \zeta(z) = 1$. $\Phi_{\rm a}(R)$ and $\Sigma_{\rm a}(R)$ are the amplitudes of the potential and the surface density, respectively. $m=2$ is the spiral's azimuthal wave number, $\Omegap=28 \Gyr^{-1}$ is its pattern speed \citep[e.g.,][]{Grosbol2018spiral,Dias2019spiral,Monteiro2021distribution}, and $\alpha = 12^\circ$ is its pitch angle \citep[e.g.,][]{Vallee2015Different,Eilers2020Spiral}. The radial wave number is $k(R) = m \cot \alpha / R \sim 1.5 \kpc^{-1}$ at $R = 6.2 \kpc$. The potential-density pair satisfies the Poisson's equation $\nabla^2\Phi_1 = 4 \pi G \rho_1$. We compute the Poisson's equation using the tight-winding approximation, i.e., $kR \gg 1$ and $k |\drm \ln \Phi_{\rm a}/\drm R|^{-1} \gg 1$. In this regime, the Poisson equation reduces to
\begin{align}
  \left[ \pd^2_z \xi(z) - k(R)^2 \xi(z) \right] \Phi_{\rm a}(R) = 4 \pi G \zeta(z) \Sigma_{\rm a}(R).
  \label{eq:spiral_poisson}
\end{align}
In a razor thin disk, i.e., $\zeta(z) = \delta(z)$, the vertical profile of the potential is exponential $\xi(z) = - \e^{-k|z|}$ \citep{binney2008galactic}, which generates a discontinuous force at the mid-plane. In a disk with finite thickness, we expect the potential to be smooth such that the force continuously switches sign at the origin. We therefore consider the following softened exponential
\begin{align}
  \xi(z) = - \e^{- k \sqrt{z^2+\zs^2}},
  \label{eq:spiral_xi}
\end{align}
where $\zs$ is the softening length. This functional form has an advantage over commonly used functions such as $\sech^2(kz)$ \citep[e.g.,][]{Faure2014Radial} in that the degree of softening can be freely adjusted. Substituting this to the Poisson's equation yields 
\begin{align}
  \zeta(z) = \frac{\Phi_{\rm a} k^2}{4 \pi G \Sigma_{\rm a}} \left(\frac{\zs^2}{z^2+\zs^2}\right) \left(1+\frac{1}{k\sqrt{z^2+\zs^2}}\right) \e^{- k \sqrt{z^2+\zs^2}}.
  \label{eq:zeta}
\end{align}
Since $\zeta>0$ at $z=0$, $\zeta=0$ at $|z| \rightarrow \infty$, and $\zeta'<0~(\zeta'>0)$ at $z>0~(z<0)$, $\zeta$ is nowhere negative. We next determine $\Phi_{\rm a}(R)$ from the normalization condition:
\begin{align}
  1 =& \int_{-\infty}^{\infty} \drm z \zeta(z) = \frac{\Phi_{\rm a} k}{2 \pi G \Sigma_{\rm a}} P(a),
  \label{eq:normalization}
\end{align}
where
\begin{align}
  P(a) = \int_{0}^{\infty} \drm x \frac{a^2}{x^2+a^2} \left(1+\frac{1}{\sqrt{x^2+a^2}}\right) \e^{- \sqrt{x^2+a^2}},
  \label{eq:Pa}
\end{align}
and $x \equiv kz$ and $a \equiv k\zs$. To compute the integral, we first consider the following function
\begin{align}
  I(a) = \int_{0}^{\infty} \drm x \e^{- \sqrt{x^2+a^2}} = a K_1(a),
  \label{eq:Ia}
\end{align}
where $K_1(a)$ is the modified Bessel function of the second kind. Equation (\ref{eq:Ia}) is obtained by changing variables $x=a\sinh(t)$. Taking the second derivative of $I$ with respect to $a$, we find
\begin{align}
   I''(a) &= a^{-1} I'(a) + P(a), \\
   \therefore P(a) &= a K_1''(a) + K_1'(a) - a^{-1} K_1(a).
   \label{eq:Ia_derivative}
\end{align}
Using the relation $K_{\nu}' = - K_{\nu-1} - \nu a^{-1} K_{\nu}$ and $K_{-\nu}=K_{\nu}$, we get
\begin{align}
  P(a) = I(a) = a K_1(a).
  \label{eq:Pa_final}
\end{align}
Hence, from equation (\ref{eq:normalization}), we finally have
\begin{align}
  \Phi_{\rm a}(R) = \frac{2 \pi G \Sigma_{\rm a}(R)}{k(R)} a K_1(a).
  \label{eq:Phia}
\end{align}
In the limit $\zs \rightarrow 0 ~(a \rightarrow 0)$, the equation recovers the solution for the razor thin disk, i.e., $\Phi_{\rm a} = 2 \pi G \Sigma_{\rm a}/k$.

We assume that the amplitude of the spiral's surface density has a Gaussian profile with a peak at the corotation radius $\RCR \equiv \vc / \Omegap$:
\begin{align}
  \Sigma_{\rm a}(R) = \Sigma_{\rm max} \e^{-(R-\RCR)^2/(2R_\beta^2)},
  \label{eq:spiral_potential_Sigmaa}
\end{align}
where $R_\beta \equiv \beta \sqrt{2}\RCR/m$ is the width of the radial profile, which is controlled by the parameter $\beta$ defined as the ratio between $R_\beta$ and the distance between the corotation resonance and the Lindblad resonances, $\sqrt{2}\RCR/m$ \citep{Hamilton2024Why}. 
Note that our model differs from that of \cite{Hamilton2024Why}, who assume a Gaussian profile for the potential, rather than for the surface density. The potential amplitude of our model is
\begin{align}
  \Phi_{\rm a}(R) = \frac{2 \pi G \Sigma_{\rm max} a K_1(a)}{m \cot\alpha} R \e^{-(R-\RCR)^2/(2R_\beta^2)},
  \label{eq:Phia}
\end{align}
which similarly decays away from the corotation radius, although the peak is shifted to
\begin{align}
  R_{\rm peak} = \RCR \frac{1 + \sqrt{1 + 8 \beta/m^2}}{2}.
  \label{eq:Phia}
\end{align}
In our model with $m=2$ and $\beta=0.5$, we have $R_{\rm peak} \sim 1.2 \RCR$.

\section{Choice of coordinate system for small-scale kicks}
\label{sec:app_diffusion}

\setcounter{figure}{0}
\renewcommand{\thefigure}{C\arabic{figure}}

\begin{figure*}
  \begin{center}
    \includegraphics[width=17.6cm]{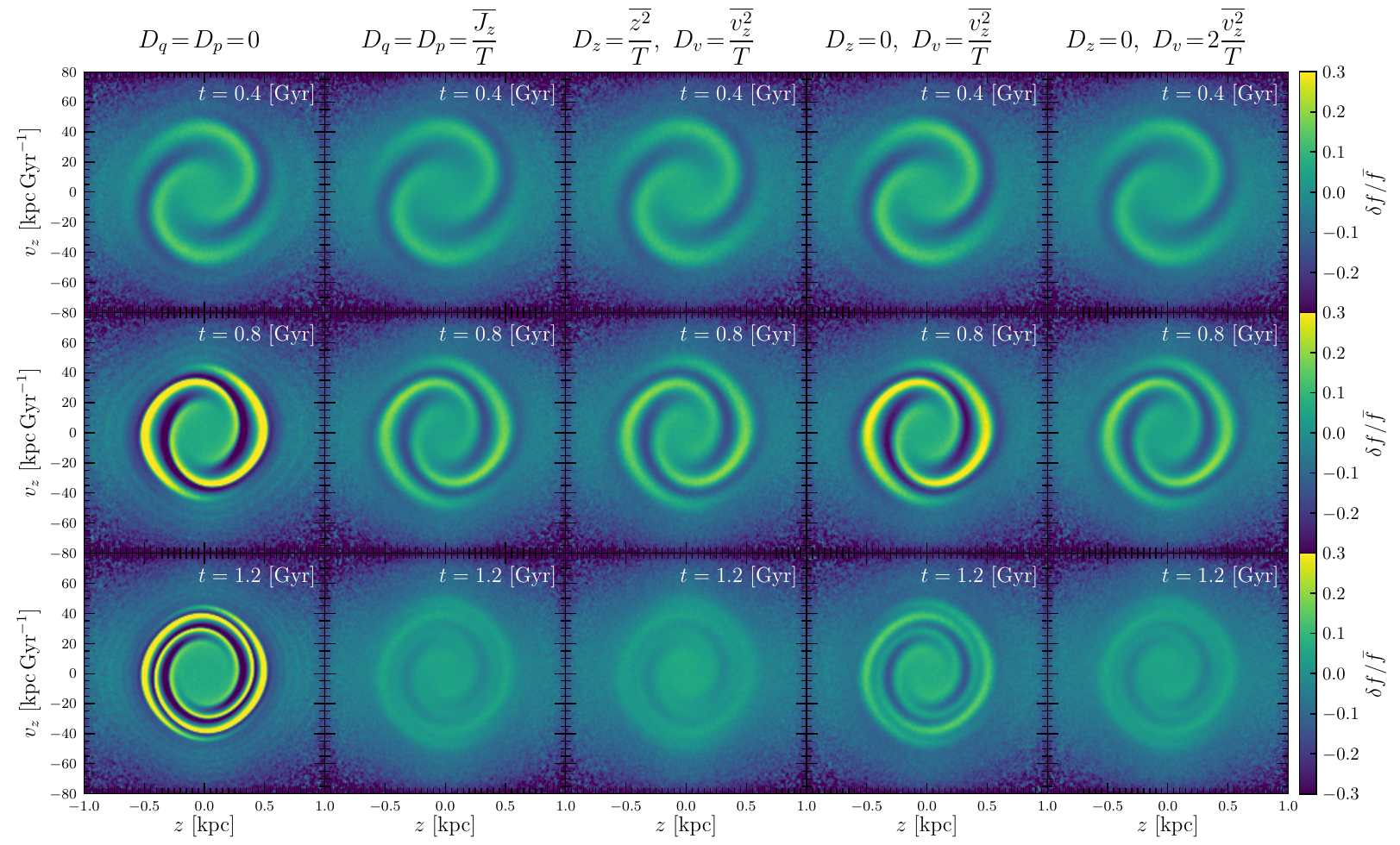}
    \caption{Comparison of transient-spiral simulations with small-scale stochastic kicks applied in different coordinates. From left to right: no kicks, kicks in $(q,p)$, kicks in $(z,v_z)$, kicks in $v_z$ only, and kicks in $v_z$ only but twice the strength. The result depends little on whether stars undergo random walk in $(q,p)$ or $(z,v_z)$. The result also depends little on whether stars are kicked in both $z$ and $v_z$ or only in $v_z$, if the diffusion coefficient in $v_z$ is doubled such that the resulting change in the mean action remains approximately the same.}
    \label{fig:D}
  \end{center}
\end{figure*}

In our kinetic analyses (Sections \ref{sec:linear_theory} and \ref{sec:nonlinear_theory}), we modeled the effect of small-scale stochastic kicks from, e.g., giant molecular clouds, as a random walk of stars in the $(q,p)$ space. Meanwhile, in our test-particle simulations (Section \ref{sec:simulation}), we applied the stochastic kicks in the $(z,v_z)$ space for the sake of computational efficiency. In the epicycle limit, the two coordinate systems are equivalent up to a dimensional constant. However, they are generally not identical, so we must check if subjecting stars to random walk in $(q,p)$ or $(z,v_z)$ leads to any significant differences.

The second and third columns of Fig.\,\ref{fig:D} compare test-particle simulations with small-scale kicks applied in $(q,p)$ and $(z,v_z)$. In the former, we transform coordinates from $(z,v_z)$ to $(q,p)$ every $\Delta t = 10 \Myr$, add random kicks $(\Delta q, \Delta p)$ drawn from a Gaussian distribution with zero mean and variance $D \Delta t$ (section \ref{sec:model_stochastic_perturbation}), and then transform back to $(z,v_z)$ for further integration. In the latter, we directly apply Gaussian-random kicks in $(z,v_z)$ with zero mean and variance $(D_z \Delta t, D_v \Delta t)$, where the diffusion coefficients are determined from the present mean-squared height and velocity dispersion (section \ref{sec:model_stochastic_perturbation}). We employ the transient spiral model (Section \ref{sec:transient_spiral_arm}) in order to highlight the effect of diffusion, which becomes prominent once the spiral arm has decayed. Comparison between the second and third columns confirms that whether stars are kicked in $(q,p)$ or $(z,\vz)$ makes negligible difference.

Throughout the paper, we applied stochastic kicks in both $q$ and $p$ (or both $z$ and $v_z$). However, encounters with, e.g., molecular clouds, are expected to be local, suggesting that only their velocity is affected. Thus, it may be more appropriate to scatter stars in momentum space rather than in the full phase space. In what follows, we show that this results in little difference so long as the rate at which the mean action of the disk changes is kept the same.

Let us first examine the rate of change in the mean action when stars undergo random walk in both $p$ and $q$, or only in $p$. Let $\langle \Delta q \rangle$ denote the ensemble average of the change in $q$ over time $\Delta t$, with an analogous expression for $p$. We assume that
\begin{align}
  \langle \Delta q \rangle = 
  \langle \Delta p \rangle = 
  \langle \Delta q \Delta p \rangle = 0, ~~
  \langle (\Delta q)^2 \rangle = D_q \Delta t, ~~ \langle (\Delta p)^2 \rangle = D_p \Delta t,
  \label{eq:app_dqsq_dpsq}
\end{align}
for some constants $D_q$ and $D_p$. For many uncorrelated weak kicks, the collision operator for a DF $f(\vw,t)$ takes the Fokker-Planck form \citep{binney2008galactic}
\begin{align}
  C[f] = - \frac{\pd}{\pd \vw} \cdot \left[ {\bm B}(\vw) f(\vw,t) - \frac{1}{2} \frac{\pd}{\pd \vw} \cdot \left[\mathsf{D}(\vw) f(\vw,t)\right]\right],
  \label{eq:app_FokkerPlank}
\end{align}
where
\begin{align}
  B_i(\vw) &= \int (\drm \Delta \vw) P(\vw, \Delta \vw) \Delta w_i = \frac{\langle \Delta w_i \rangle}{\Delta t}, \\
  \mathsf{D}_{ij}(\vw) &= \int (\drm \Delta \vw) P(\vw, \Delta \vw) \Delta w_i \Delta w_j = \frac{\langle \Delta w_i \Delta w_j \rangle}{\Delta t}, 
  \label{eq:app_FokkerPlank_BD}
\end{align}
with $P(\vw, \Delta \vw)$ denoting the transition probability that a star at $\vw$ is scattered by $\Delta \vw$ over time $\Delta t$.
From equation (\ref{eq:app_dqsq_dpsq}), it follows that 
\begin{align}
  C[f] = \frac{1}{2}\left( D_q \frac{\pd^2 f}{\pd q^2} + D_p \frac{\pd^2 f}{\pd p^2} \right).
  \label{eq:app_collision_operator_qp}
\end{align}
Transforming to the angle-action coordinates (equation \ref{eq:cartesian}), we have
\begin{align}
  &C[f] = \frac{D_q + D_p}{2}\frac{\pd f}{\pd \Jz} 
  + \Jz \left( D_q \sin^2\theta_z + D_p \cos^2\theta_z\right) \frac{\pd^2 f}{\pd \Jz^2} \nonumber \\
  &- \frac{\cos\thetaz\sin\thetaz}{2\Jz} \left(D_q - D_p\right) \frac{\pd f}{\pd \thetaz}
  + \frac{D_q \cos^2\theta_z + D_p \sin^2\theta_z}{4\Jz} \frac{\pd^2 f}{\pd \thetaz^2} \nonumber \\
  &+ \cos\thetaz\sin\thetaz \left(D_q - D_p\right) \frac{\pd^2 f}{\pd \Jz \pd \thetaz}.
  \label{eq:app_collision_operator_AA}
\end{align}
When $D_q=D_p=D$, the equation reduces to equation (\ref{eq:collision_operator}).

We now use this collision operator to calculate the rate at which the mean action changes. We have
\begin{align}
  \frac{\drm \overline{\Jz}}{\drm t} 
  = \frac{\drm}{\drm t} \int \drm \thetaz \int \drm \Jz \Jz f
  = \int \drm \thetaz \int \drm \Jz \Jz C[f].
  \label{eq:app_change_in_mean_action}
\end{align}
Substituting equation (\ref{eq:app_collision_operator_AA}) and integrating by parts, we find, after a tedious but straightforward calculation,
\begin{align}
  \frac{\drm \overline{\Jz}}{\drm t} = \frac{D_q + D_p}{2}.
  \label{eq:app_change_in_mean_action_final}
\end{align}
The result suggests that the change in the mean action in case $(D_q,D_p)=(0,D)$ is exactly half of that in case $(D_q,D_p)=(D,D)$. In other words, if we set $(D_q,D_p)=(0,2D)$, the rate of change in the mean action would be the same as $(D_q,D_p)=(D,D)$. Since the typical orbital period ($T_z = 2\pi/\Omegaz \sim 0.1 \Gyr$) is much shorter than the diffusion timescale, which is of order the age of the disk ($T \sim 10 \Gyr$), equation (\ref{eq:app_change_in_mean_action_final}) implies that as long as the diffusion coefficients are determined based on the present mean action of the disk (i.e., $D_q+D_p$ remains constant), the choice of kicking in $q$ and $p$, or only in $p$ has little effect on the evolution of $f$.

It is difficult to derive an analytical expression for $\drm \overline{\Jz}/\drm t$ when stars undergo random walks in the $(z,v_z)$ coordinates. However, we expect a similar conclusion, since kicks in $(z,v_z)$ and $(q,p)$ result in little difference, as demonstrated earlier (second and third columns of Fig.\,\ref{fig:D}). We verify this numerically by running simulations where stars are only kicked in $v_z$ (forth and fifth columns of Fig.\,\ref{fig:D}). Compared to the standard model, where stars are kicked in both $z$ and $v_z$ (3rd column), removing kicks in $z$ (4th column) results in a reduced diffusion. However, when the strength of kick in $v_z$ is increased by a factor of two (5th column), the degree of diffusion is almost the same as the standard model (3rd column). These results confirm that the consequence of small-scale kicks depends little on whether stars are kicked isotropically in phase space or only in $v_z$, provided the strength of kick in $v_z$ is doubled, keeping the rate of change in the mean action roughly the same.

\section{Solution to the linearized kinetic equation}
\label{sec:app_solution_LKE}

This appendix details the derivation of the solution to the linearized kinetic equation (\ref{eq:kinetic_eq_lin_Fourier}):
\begin{align}
  \frac{\pd \hat{f}_{\nz}}{\pd t} + i \nz \Omegaz \hat{f}_{\nz} - i \nz \frac{\pd f_0}{\pd \Jz} \hat{\Phi}_{\nz} = D \Jz \frac{\pd^2 \hat{f}_{\nz}}{\pd \Jz^2}.
  \label{eq:app_kinetic_eq_lin_Fourier}
\end{align}
Following \cite{Tremaine2023Snail}, we seek an approximate solution by Taylor expanding the equation around a reference action $J_{z0}$, i.e., $\Jz = J_{z0} + \Delta \Jz$ and $\Omegaz = \Omega_{z0} + \Omega'_{z0} \Delta \Jz$. Then the equation becomes
\begin{align}
  \frac{\pd \hat{f}_{\nz}}{\pd t} + i \nz (\Omega_{z0} + \Omega'_{z0} \Delta \Jz) \hat{f}_{\nz} - D (J_{z0} + \Delta \Jz) \frac{\pd^2 \hat{f}_{\nz}}{\pd (\Delta\Jz)^2} = g(J_{z0}, t),
  \label{eq:app_kinetic_eq_lin_Fourier_Taylor}
\end{align}
where
\begin{align}
  g(\Jz,t) = i \nz \frac{\pd f_0}{\pd \Jz} \hat{\Phi}_{\nz}.
  \label{eq:app_g}
\end{align}
The order of the last term in the left hand side of equation (\ref{eq:app_kinetic_eq_lin_Fourier_Taylor}) compared to the third term is 
\begin{align}
  \frac{D \Delta \Jz \pd^2 \hat{f}_{\nz}/\pd (\Delta\Jz)^2}{\nz \Omega'_{z0} \Delta \Jz \hat{f}_{\nz}}
  \sim \frac{D}{\nz \Omega'_{z0} J_{\rm h}^2},
  \label{eq:app_order_last_term_LHS}
\end{align}
where $J_{\rm h}$ is the scale over which $\hat{f}_{\nz}$ varies. It is reasonable to associate $J_{\rm h}$ with the half-width of the resonant island in $\Jz$ space, $J_h = 2 \sqrt{-|\hat{\Psi}_{\nz}|/\Omega'_{z0}}$ \citep[e.g.,][]{Hamilton2023BarResonanceWithDiffusion}. Using equation (\ref{eq:dimensionless_diff_param}), it follows that the ratio of the two terms scales as $\epsilon^{1/2}\Delta/4$, where $\epsilon \sim \hat{\Psi}_{\nz}/H_0$. Since $\epsilon \sim 0.1$ and $\Delta \sim 0.2$ in our model, we will neglect the last term in the left hand side of equation (\ref{eq:app_kinetic_eq_lin_Fourier_Taylor}):
\begin{align}
  \frac{\pd \hat{f}_{\nz}}{\pd t} + i \nz (\Omega_{z0} + \Omega'_{z0} \Delta \Jz) \hat{f}_{\nz} - D J_{z0} \frac{\pd^2 \hat{f}_{\nz}}{\pd (\Delta\Jz)^2} = g(J_{z0}, t).
  \label{eq:app_kinetic_eq_lin_Fourier_Taylor_IgnoreLastTerm}
\end{align}
We caution though that, strictly speaking, linear theory is only valid for $\Delta \gtrsim 1$ \citep[see][]{Hamilton2023BarResonanceWithDiffusion}. 

\setcounter{figure}{0}
\renewcommand{\thefigure}{E\arabic{figure}}

\begin{figure*}
  \begin{center}
    \includegraphics[width=15.9cm]{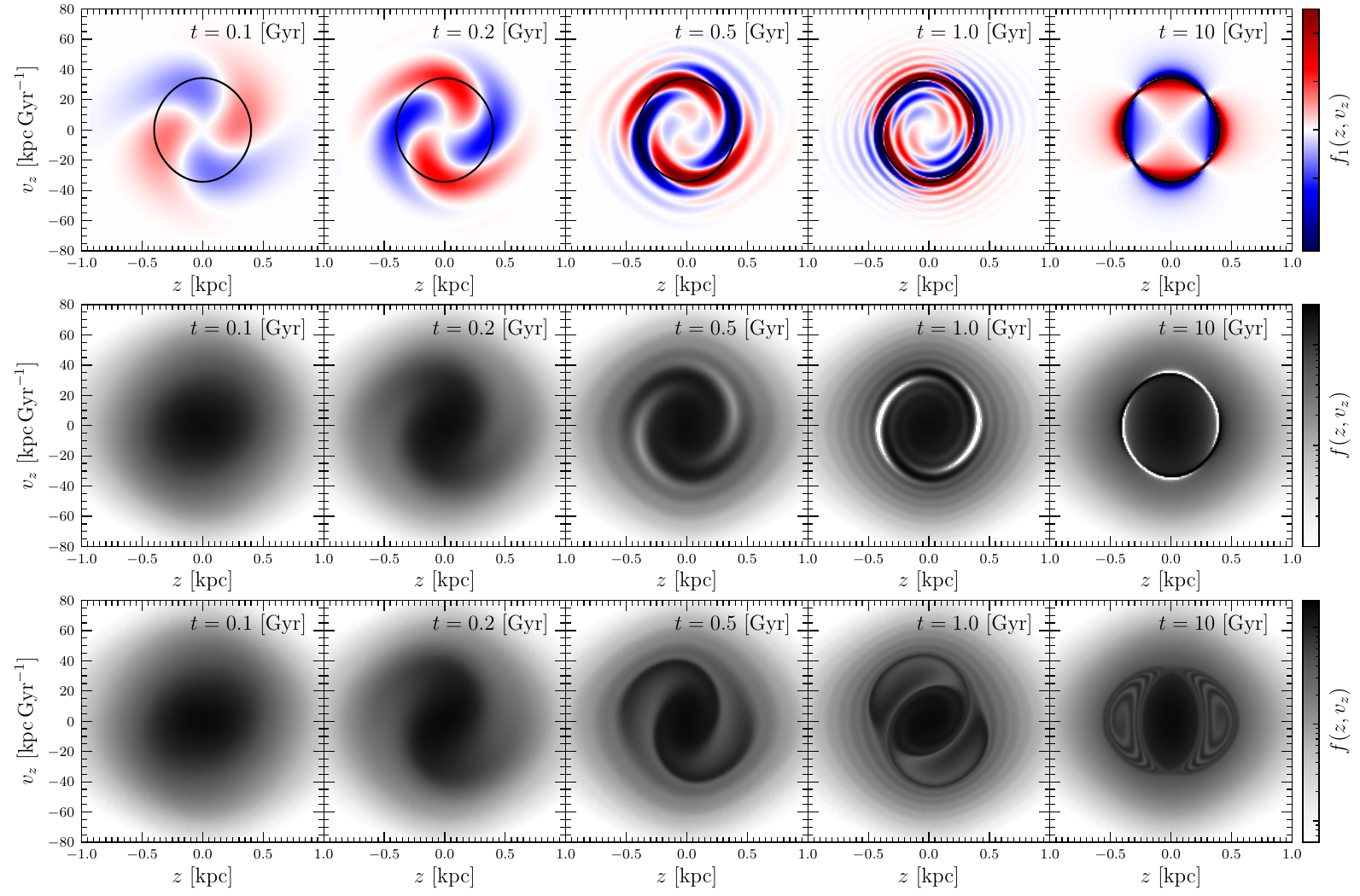}
    \caption{Disk response against a persistent resonant perturbation in the collisionless case at different times increasing from left to right, calculated using linear theory (top and middle) and test-particle simulation (bottom). The linear response becomes increasingly inaccurate near the resonance (black curve), where stars become trapped. Small-scale diffusion erases these erroneous structures and keeps the linear response well-behaved at the resonance (see Fig.\,\ref{fig:linear_zvz}).}
    \label{fig:linear_zvz_NoCol}
  \end{center}
\end{figure*}

We look for a solution with the initial condition $\hat{f}_{\nz}(\Jz,t=0)=0$ by Laplace transforming the equation in time. Multiplying by $\e^{-pt}$ and integrating over $t \in [0,\infty]$, we have
\begin{align}
  [p + i \nz (\Omega_{z0} + \Omega'_{z0} \Delta \Jz)] \tilde{f}_{\nz} - D J_{z0} \frac{\pd^2 \tilde{f}_{\nz}}{\pd (\Delta\Jz)^2} = \tilde{g}(J_{z0},p),
  \label{eq:app_kinetic_eq_lin_Fourier_Taylor_Laplace}
\end{align}
where the tilde denotes functions in Laplace space, $\tilde{f}_{\nz}(\Jz,p) = \int_0^\infty \drm t \e^{-pt} \hat{f}_{\nz}(\Jz,t)$, and is similarly defined for $\tilde{g}$. To simplify the equation, we define a new dimensionless variable
\begin{align}
  x \equiv - t_0 [p + i \nz (\Omega_{z0} + \Omega'_{z0} \Delta \Jz)],
  \label{eq:app_x}
\end{align}
where $t_0$ will be determined later. The equation then reads 
\begin{align}
  -x \tilde{f}_{\nz} + \nz^2 {\Omega'_{z0}}^2 D J_{z0} t_0^3 \frac{\pd^2 \tilde{f}_{\nz}}{\pd x^2} = t_0 \tilde{g},
  \label{eq:app_kinetic_eq_lin_Fourier_Laplace_Taylor_x}
\end{align}
We now choose $t_0$ to be $t_0 \equiv (\nz^2 {\Omega'_{z0}}^2 D J_{z0})^{-1/3}$ such that 
\begin{align}
  -x \tilde{f}_{\nz} + \frac{\pd^2 \tilde{f}_{\nz}}{\pd x^2} = t_0 \tilde{g},
  \label{eq:app_kinetic_eq_lin_Fourier_Laplace_Taylor_x_2}
\end{align}
which is an inhomogeneous Airy equation. In the absence of the potential perturbation ($\tilde{g}=0$), the solution is given by the Airy function \citep[e.g.,][]{Tremaine2023Snail}. For the general inhomogeneous case, the solution is given by the Scorer function \citep{scorer1950numerical,Gil2001nonoscillating}:
\begin{align}
  \tilde{f}_{\nz} = t_0 \tilde{g} \pi Hi(x), ~~~ {\rm where} ~~~ Hi(x) = \frac{1}{\pi} \int_0^\infty \drm s \e^{xs - s^3/3}.
  \label{eq:app_kinetic_eq_lin_Fourier_Laplace_Taylor_x_2}
\end{align}
To obtain $\hat{f}_{\nz}$, we substitute $x$ (\ref{eq:app_x}) back into the equation and apply the inverse Laplace transform:
\begin{align}
  \hat{f}_{\nz} 
  &= \mathcal{L}^{-1}[\tilde{f}_{\nz}] \nonumber \\
  &= t_0 \!\int_0^\infty\! \drm s \e^{-i\nz (\Omega_{z0} + \Omega'_{z0} \Delta \Jz) t_0 s - s^3/3} \mathcal{L}^{-1}[\tilde{g}(p) \e^{-t_0 s p}] \nonumber \\
  &= t_0 \!\int_0^\infty\! \drm s \e^{-i\nz (\Omega_{z0} + \Omega'_{z0} \Delta \Jz) t_0 s - s^3/3} \! \!\int_0^t \!\drm t' g(t') \delta(t \!-\! t_0 s \!-\! t').
  \label{eq:app_fn_inverse_Laplace}
\end{align}
Performing the integral over $s$ using the delta function, we get
\begin{align}
  \hat{f}_{\nz}(J_{z0} \!+\! \Delta \Jz, t) \!=\! \!\int_0^t\! \!\drm t'\! \e^{-i\nz(\Omega_{z0} + \Omega'_{z0} \Delta \Jz) (t-t') - [(t-t')/\td]^3} \!g(J_{z0}, t'),
  \label{eq:app_fn_solution}
\end{align}
where $\td \equiv 3^{1/3}t_0$ is the diffusion timescale (equation \ref{eq:diffusion_timescale}). It can be verified that the above solution satisfies equation (\ref{eq:app_kinetic_eq_lin_Fourier_Taylor_IgnoreLastTerm}).
Finally, taking the limit $\Delta \Jz \rightarrow 0$, we obtain the following approximate solution (equation \ref{eq:kinetic_eq_lin_sol_fn})
\begin{align}
  \hat{f}_{\nz}(J_{z0}, t) \simeq \int_0^t \drm t' \e^{-i\nz \Omega_{z0} (t-t') - [(t-t')/\td]^3} g(J_{z0}, t').
  \label{eq:app_fn_solution_limitDJz0}
\end{align}
In the limit of zero diffusion $\td \rightarrow \infty$, we recover the solution to the linearized collisionless Boltzmann equation \citep[e.g.,][]{Kalnajs1971Dynamics}:
\begin{align}
  \hat{f}_{\nz}(J_{z}, t) = \int_0^t \drm t' \e^{-i\nz \Omega_{z} (t-t') } g(J_{z}, t').
  \label{eq:app_fn_solution_collisionless}
\end{align}

\section{Linear response in the collisionless limit}
\label{sec:app_linear_response_collisionless}

In the collisionless limit, the linear approximation becomes invalid near the resonance at late times as discussed in Section \ref{sec:nonlinear_theory}. This Appendix demonstrates this using equation (\ref{eq:app_fn_solution_collisionless}). Substituting equations (\ref{eq:app_g}) and (\ref{eq:spiral_potential_Fourier_coefficient}) to (\ref{eq:app_fn_solution_collisionless}), and performing the inverse Fourier transform (\ref{eq:Fourier_expand_f1_Phi1}) under the assumption that the perturbation is introduced instantaneously at $t=0$ (i.e., $\mathcal{T}=1$), we obtain \citep{Weinberg2007BarHaloInteraction,Chiba2022Oscillating}
\begin{align}
  f_1(\thetaz, \Jz, t) = \frac{1}{2} \sum_{\nz,l} \nz \frac{\pd f_0}{\pd J_z} |\hat{\Psi}_{\nz}| \sin(\thetas - \Omegas t/2) \frac{\sin(\Omegas t/2)}{\Omegas/2},
  \label{eq:app_fn_solution_collisionless_const}
\end{align}
where $\thetas = \nz \thetaz - l \omega t + \arg \hat{\Psi}_{\nz}$ and $\Omegas = \nz \Omegaz - l \omega$. Note the factor $\sin(\Omegas t/2)/(\Omegas/2)$, which implies that the linear response at the resonance grows linearly with time\footnote{A common misconception is that linear response diverges on resonance. However, equations (\ref{eq:app_fn_solution_collisionless_const}-\ref{eq:app_fn_solution_collisionless_sinc}) demonstrate that the linear response is everywhere regular at all finite times and divergence only occurs in the time-asymptotic limit $t \rightarrow \infty$. A closely related misconception is that, in the self-gravitating case, linear response diverges for resonantly-driven systems at marginal stability. Again this is also only true as $t \rightarrow \infty$ \citep{Hamilton2023marginalstability}.}:
\begin{align}
  \lim_{\Omegas\rightarrow0}\frac{\sin(\Omegas t/2)}{\Omegas/2}=t.
  \label{eq:app_fn_solution_collisionless_sinc}
\end{align}

Fig.\,\ref{fig:linear_zvz_NoCol} illustrates the breakdown of the linear approximation in the collisionless limit. The top, middle, and bottom rows plot the linear perturbation $f_1$, the full linear response $f=f_0+f_1$, and the full nonlinear response from a test-particle simulation, respectively. At early times, linear theory accurately captures the growth of the phase spiral, but at late times it deviates from the nonlinear response, since the former grows linearly with time at the resonance, while the latter saturates due to orbital trapping. In the time-asymptotic limit, $t \rightarrow \infty$, the linear response in the collisionless case diverges at the resonance. By contrast, in the `collisional' case (Section \ref{sec:linear_theory}), the small-scale random kicks suppress the development of these spurious structures, allowing the linear response to remain well-behaved at all times (see Fig.\,\ref{fig:linear_zvz}).


\bsp	
\label{lastpage}
\end{document}